\documentclass[a4paper,fleqn,usenatbib]{mnras}

\usepackage[T1]{fontenc}
\usepackage{ae,aecompl}

\usepackage{amssymb}	
\usepackage{amsmath}
\usepackage{graphicx}
\usepackage[usenames]{color}
\usepackage{bm} 

\usepackage{xspace}

\usepackage{multirow}

\usepackage{lineno}

\newcommand{\Ddel}{\delta_{\rm D}   }
\newcommand{\MpcOh}{ \,  \mathrm{Mpc}  \, h^{-1} }

\newcommand{\nn}{ \nonumber }

\newcommand{\xip}{ \xi_{\rm p}    }

\newcommand{\beq}{\begin{equation}}
\newcommand{\eeq}{\end{equation}}
\newcommand{\beqa}{\begin{eqnarray}}
\newcommand{\eeqa}{\end{eqnarray}}

\newcommand\T{\rule{0pt}{2.6ex}}       
\newcommand\B{\rule[-1.2ex]{0pt}{0pt}} 

\newcommand{\change}[1]{{\textcolor{black}{#1}}\xspace}

\newif\ifcomments
\commentstrue   

\usepackage{eso-pic}

\AddToShipoutPictureBG*{%
  \AtPageUpperLeft{%
    \hspace{0.75\paperwidth}%
    \raisebox{-3.5\baselineskip}{%
      \makebox[0pt][l]{\textnormal{FERMILAB-PUB-21-490-AE}}
}}}%

\AddToShipoutPictureBG*{%
  \AtPageUpperLeft{%
    \hspace{0.75\paperwidth}%
    \raisebox{-4.5\baselineskip}{%
      \makebox[0pt][l]{\textnormal{DES-2021-0667}}
}}}%

\title[ $\xi_{\rm p}$ with general photo-$z$ uncertainties  ]{ Clustering with general photo-$z$ uncertainties:  
  Application to Baryon Acoustic Oscillations }

\author[authors]{authors}

\author[K.~C.~Chan, et al]{
\parbox{\textwidth}{
Kwan Chuen Chan$^{1,2}$\thanks{E-mail: chankc@mail.sysu.edu.cn (KCC)},
Ismael Ferrero$^{3,4,5}$,
Santiago Avila$^{6,7}$,  
Ashley J.~Ross$^{8}$,   \\
Martin Crocce$^{4,5}$,
Enrique Gazta\~naga$^{4,5}$
}  \\  \\
$^{1}$ School of Physics and Astronomy, Sun Yat-sen University, 2 Daxue Road, Tangjia, Zhuhai, 519082, China  \\
$^{2}$ CSST Science Center for the Guangdong-Hongkong-Macau Greater Bay Area, SYSU, China \\
$^{3}$ Institute of Theoretical Astrophysics, University of Oslo. P.O. Box 1029 Blindern, NO-0315 Oslo, Norway    \\
$^{4}$ Institut d'Estudis Espacials de Catalunya (IEEC), 08034 Barcelona, Spain  \\
$^{5}$ Institute of Space Sciences (ICE, CSIC),  Campus UAB, Carrer de Can Magrans, s/n,  08193 Barcelona, Spain  \\ 
$^{6}$ Instituto de Fisica Teorica UAM/CSIC, Universidad Autonoma de Madrid, 28049 Madrid, Spain    \\
$^{7}$ Departamento de F\'isica Te\'orica,  Universidad Aut\'onoma de Madrid, 28049 Madrid, Spain \\ 
$^{8}$ Center for Cosmology and Astro-Particle Physics, The Ohio State University, Columbus, OH 43210, USA \\
}

\date{Accepted XXX. Received YYY; in original form ZZZ}

\pubyear{AAAA}

\begin{document}
\label{firstpage}
\pagerange{\pageref{firstpage}--\pageref{lastpage}}
\maketitle

\begin{abstract}

  Photometric data can be analyzed using the three-dimensional correlation function $\xi_{\rm p}$ to extract cosmological information via e.g., measurement of the Baryon Acoustic Oscillations (BAO). Previous studies modeled $\xi_{\rm p} $  assuming a Gaussian photo-$z$ approximation. In this work we improve the modeling by incorporating realistic photo-$z$ distribution.  We show that the position of the BAO scale in $\xi_{\rm p}$ is determined by the photo-$z$ distribution and the Jacobian of the transformation. The latter diverges at the transverse scale of the separation $s_\perp $, and it explains why $\xi_{\rm p } $ traces the underlying correlation function at $s_\perp $, rather than $s$, when the photo-$z$ uncertainty $ \sigma_z / (1+ z) \gtrsim 0.02$.   We also obtain the Gaussian covariance for $\xi_{\rm p}$.  Due to photo-$z$ mixing, the covariance of $\xi_{\mathrm{p}}$ shows strong off-diagonal elements.  The high correlation of the data causes some issues to the data fitting. Nonetheless, we find that either it can be solved by suppressing the largest eigenvalues of the covariance or it is not directly related to the BAO.  We test our BAO fitting pipeline using a set of mock catalogs. The data set is dedicated for Dark Energy Survey Year 3 (DES Y3) BAO analyses and includes realistic photo-$z$ distributions. The theory template is in good agreement with mock measurement.  Based on the DES Y3 mocks, $\xi_{\rm p}$ statistic is forecast to constrain the BAO shift parameter $\alpha$ to be $1.001 \pm 0.023$, which is well consistent with the corresponding constraint derived from the angular correlation function measurements. Thus $\xi_{\rm p}$ offers a competitive alternative for the photometric data analyses.

\end{abstract}

\begin{keywords}
  cosmology: observations - (cosmology:) large-scale structure of Universe
\end{keywords}

\maketitle

\section{Introduction} 
Imaging surveys infer the redshift of galaxy samples, photo-$z$, using broadband filters. Ongoing and future large-scale structure surveys that collect enormous amount  of photo-$z$ data include Kilo-Degree Survey (KiDS) \footnote{http://kids.strw.leidenuniv.nl}, Dark Energy Survey (DES) \footnote{https://www.darkenergysurvey.org},  Hyper Suprime-Cam (HSC) \footnote{https://www.naoj.org/Projects/HSC}, Rubin Observatory's Legacy Survey of Space and Time (LSST) \footnote{https://www.lsst.org}, Euclid \footnote{https://www.euclid-ec.org}, and the Chinese Survey Space Telescope (CSST) \footnote{http://www.nao.cas.cn/csst}.  These surveys typically use several bandpasses, e.g.~DES has $grizY$ and the accuracy of the resultant photo-$z$ is about $\sigma \sim 0.03(1+z)$ [see e.g.~\citet{Crocce_etal2019,y3-baosample}, but it can be improved if more stringent criteria are applied \citep{Rozo_etal2016}]. Although photometric surveys yield less precise redshift than spectroscopic ones ($\sigma \sim 10^{-3}$), they are more efficient to collect a large volume of survey data with deep magnitude.

Baryonic Acoustic Oscillations (BAO) \citep{PeeblesYu1970,SunyaevZeldovich1970} are the primordial acoustic features imprinted  in the distribution of the large-scale structure.   In the early universe, photons and baryons form a tightly coupled plasma, and acoustic oscillations are excited in it. After the recombination, hydrogen atoms form, the acoustic waves get stalled because there is no medium for propagation.  The acoustic patterns are preserved in the distribution of galaxies and they correspond to the sound horizon at the drag epoch, which is about 150 Mpc. Formation of the BAO is governed by  well-understood linear physics,  see \citet{BondEfstathiou1984,BondEfstathiou1987,HuSugiyama1996,HuSugiyamaSilk1997,Dodelson_2003} for the details of the cosmic microwave background physics. Given its robustness, BAO has been regarded as a standard ruler in cosmolology, see e.g.~\cite{WeinbergMortonson_etal2013,Aubourg:2014yra}.   BAO has been measured in numerous spectroscopic data analyses. It was first clearly detected by \citet{Eisenstein_etal2005, Cole_etal2005} and followed up in many subsequent studies, e.g.~\citet{Gaztanaga:2008xz, Percival_etal2010, Beutler_etal2011, Blake2012_WiggleZ, Anderson_BOSS2012, Kazin_etal2014, Ross_etal2015, Alam_etal2017, Alam:2020sor}.

The sample size in terms of volume and depth from photometric surveys may compensate its imprecision in redshift information and give competitive measurements \citep{SeoEisenstein_2003,BlakeBridle_2005,Amendola_etal2005,Benitez:2008fs,Zhan:2008jh, Chaves-Montero:2016nmw}. Indeed, there are several detections of the BAO feature in the galaxy distribution using photometric data from SDSS \citep{Padmanabhan_etal2007,EstradaSefusattiFrieman2009, Hutsi2010, Seo_etal2012,Carnero_etal2012, deSimoni_etal2013}, DES [Y1 \citep{Abbott:2017wcz} and Y3 \citep{y3-baomain}, hereafter DES Y3],  and DECaLS \citep{Sridhar_etal2020}.

Because of the photo-$z$ uncertainty, it is natural to perform the analyses of the photometric data using angular statistics such as the angular correlation function \citep{Coupon:2011mz,DES:2015jcy,Abbott:2017wcz,DES:2017hdw,Vakili:2020dwl} or the angular power spectrum \citep{Padmanabhan_etal2007,Seo_etal2012,DES:2018csk,LSST:2019wqx}.  In particular, these two statistics were adopted as the fiducial statistical tools to measure BAO in DES Y3 analyses.  However, given enormous amount photo-$z$ data available, it is worthwhile exploring new methods to enhance information extraction, especially from the BAO. \citet{Ross:2017emc} proposed to measure the BAO using the three-dimensional (3D) correlation function characterized by the transverse scale of the separation. This statistic has been applied to photo-$z$ data to get interesting BAO measurements \citep{Abbott:2017wcz,Sridhar_etal2020}. The photo-$z$ uncertainty of the sample must be folded in the theory template. To simplify the modeling,  the photo-$z$ distribution was assumed to be Gaussian in  \citet{Ross:2017emc}. The actual photo-$z$ distribution has non-negligible deviation from Gaussian distribution, e.g.~\citet{Crocce_etal2019,Zhou_etal2021,y3-baosample}. This assumption may risk introducing bias, and so this statistic was not adopted in the fiducial DES Y1 and Y3 analyses. In this work, we improve the modeling of the 3D clustering by incorporating general photo-$z$ distribution and present a Gaussian covariance for it.

This paper is organized as follows.  In Sec.~\ref{sec:theory_template}, we first review the computation of the angular correlation and then show that the general angular correlation can be turned into the 3D correlation function $\xi_{\rm p} $.  We consider the Gaussian photo-$z$ limit and use it to understand the effect of photo-$z$ uncertainties on  $\xi_{\rm p} $ in Sec.~\ref{sec:Gaussian_photoz_limit}. In Sec.~\ref{eq:Gaussian_covmat}, we derive the Gaussian covariance for   $\xi_{\rm p} $  using the Gaussian covariance for the angular correlation function. In Sec.~\ref{sec:Template_and_Covariance}, we  present the numerical results on the template and covariance, and compare them with mock results whenever possible.  We discuss the issue in data fitting caused by highly correlated data in Sec.~\ref{sec:highly_correlated_fit}, and then present the BAO fit results in Sec.~\ref{sec:BAOfit_results}. We conclude in Sec.~\ref{sec:Conclusions}.  In Appendix ~\ref{sec:specz_check}, we test our pipeline for the case of spectroscopic data.   The default cosmology adopted in this work  is  the MICE cosmology \citep{Fosalba_etal2015,Crocce_etal2015}, which is a flat $ \Lambda$CDM with $\Omega_{\rm m} = 0.25$,  $\Omega_{\Lambda} = 0.75$, $h=0.7$, and $\sigma_8 = 0.8$.

\section{ Angular and three-dimensional two-point correlation } 
\label{sec:theory_template}

In this section, we first review the computation of the angular correlation function, $w$,  in the presence of general photo-$z$ uncertainty.  We then use the general angular correlation function to derive the 3D two-point correlation $ \xi_{\rm p } $.

\subsection{Angular correlation function $w_p$ }

Because in imaging surveys, angles are well determined but there are typically significant redshift uncertainties, it is natural to use angular quantities.  It is useful to first clarify the relation between volume number density and tomographic bin angular number density.

\change{ Suppose that there are $dN$ objects in a volume element parametrized by the redshift interval $dz $ and the solid angle $d \Omega$, the volume number density $ n_{(3)}$ is given by 
\beq
\label{eq:n3_def} 
n_{(3)}( \bm{r} ) =  \frac{ dN }{ dz d \Omega }.
\eeq
The tomographic angular number density $n_{(2)} $ is defined as 
\beq
\label{eq:n2_def} 
n_{(2)}(z, \hat{\bm{r}} ) = \frac{ dN }{ d \Omega } =   n_{(3)} ( \bm{r} )  dz . 
\eeq
}
 We can go on to define the volume density contrast $ \delta_{(3)} $  and angular density contrast $ \delta_{(2)} $ 
\begin{align}
  \delta_{(3)} &= \frac{n_{(3)} - \bar{n}_{(3)}  }{ \bar{n}_{(3)} }, \\
  \delta_{(2)} &= \frac{n_{(2)} - \bar{n}_{(2)}  }{ \bar{n}_{(2)} },
\end{align}
where $ \bar{n}_{(3)} $ and $\bar{n}_{(2)}$ are their respective mean values.  From Eq.~\eqref{eq:n2_def}, it is clear that $ \delta_{(3)}= \delta_{(2)} $.



\change{
We start with the effect of the photo-$z$ on the galaxy number density:
\beq
n_{\rm p} ( z_{\rm p},  \hat{ \bm{r} }  ) = \int d z f( z| z_{\rm p}) n_{ (3) }( z,  \hat{ \bm{r} }  ), 
\eeq
where  $ f( z| z_{\rm p} )  dz $ is the probability that the true redshift of the galaxy falls within $[z, z + dz]$ given that its photo-$z$ is $z_{\rm p}$,  and $ n_{\rm p} $ denotes the number density in photo-$z$ space.  For photometric data,  $f(z|z_{\rm p} ) $ is of central importance and it can be estimated by the photo-$z$ codes using template fitting or machine learning methods [see e.g.~\citet{Benitez:1998br,Ilbert:2006dp,Sadeh:2015lsa,DeVicente:2015kyp}]. }  

\change{
At the background level, we have
\beq
\bar{n}_{\rm p} ( z_{\rm p} ) = \int d z f( z| z_{\rm p}) \bar{n}_{(3)}( z  ) ,  
\eeq
where $ \bar{n}_{\rm p} $ and $\bar{n}_{(3)}$ are the angular mean number density in the photo-$z$ and true redshift space, respectively.   The perturbation is given by
\begin{align}
  n_{\rm p} ( z_{\rm p},  \hat{ \bm{r} }  ) &= \bar{n}_{\rm p} (z_{\rm p} )  [ 1 + \delta_{\rm p}( z_{\rm p},  \hat{ \bm{r} }) ]   \nn \\
 & =  \int d z f( z| z_{\rm p}) \bar{n}_{(3) }(z)  [ 1 + \delta_{ (3) }( z,  \hat{ \bm{r} }  )] ,  
\end{align}
and we get
\beq
\delta_{\rm p} ( z_{\rm p} , \hat{ \bm{r} }  ) = \int d z   f(z|z_{\rm p} ) \frac{ \bar{n}_{(3) }(z) }{ \bar{n}_{\rm p} (z_{\rm p}  ) } \delta_{ (3) }( z, \hat{\bm{r}} ). 
\eeq
Therefore, the appropriate true redshift distribution for the density contrast is
\beq
\label{eq:delta_p}
\phi( z|z_{\rm p} ) = f ( z|z_{\rm p} )  \frac{ \bar{n}_{(3)}(z) }{ \bar{n}_{\rm p} (z_{\rm p}  ) }  .
\eeq
It is easy to see that $\phi( z|z_{\rm p} ) $ indeed furnishes a probability density. }
Here we assume that there are sufficiently many particles in the coarse-grained scale of interest so that the photo-$z$ effect can be modeled  by the distribution  $\phi$.   In this regime, the effect of the photo-$z$ is to average $\delta_{(3)} $ over $z$.


Using Eq.~\eqref{eq:delta_p}, the photo-$z$ angular correlation function $w_p$ can be written as
\begin{align}
  \label{eq:xiphoto_def} 
 &  w ( \theta,  z_{\rm p} ,z_{\rm p}' )  \equiv \langle \delta_{\rm p} (z_{\rm p}, \hat{ \bm{r}} )  \delta_{\rm p} (z_{\rm p}', \hat{ \bm{r}}' ) \rangle \nn \\
  =  & \int dz \phi(z|z_{\rm p}) \int dz' \phi(z'|z_{\rm p}')  \langle  \delta_{(3)} ( z,\hat{ \bm{r}} )  \delta_{(3)} ( z',\hat{ \bm{r}}' )    \rangle, 
\end{align}
with $\theta \equiv \cos^{-1} ( \hat{\bm{r}} \cdot \hat{\bm{r}}'  ) $.
By expressing  $ \delta_{(3)} $ in  terms of its Fourier modes, we have 
\begin{align}
  \label{eq:xiphoto_res1} 
     w ( \theta, z_{\rm p},z_{\rm p}' )  = &  \int dz \phi(z| z_{\rm p}) \int dz' \phi(z'|z_{\rm p}') \int \frac{d^3 k}{(2 \pi)^3 }  \nn \\
   & \quad  \times e^{ i \bm{k} \cdot [ \bm{r}(z) -  \bm{r}'(z') ]}      P(\bm{k}, z,z').  
\end{align}
Utilizing the variable $\bm{s} =  \bm{r} -  \bm{r}' $ and applying the Rayleigh expansion of the plane wave 
\beq
e^{ i \bm{k} \cdot \bm{s} } =  4 \pi \sum_{l=0}^\infty \sum_{m=-l}^{l} i^l j_l( ks ) Y_{lm}( \hat{ \bm{k} } )  Y_{lm}^* (\hat{ \bm{s} }  ) ,  
\eeq
we have
\begin{align}
  \label{eq:w_photoz}
    w( \theta, & z_{\rm p},z_{\rm p}' )
  =   \int dz \phi(z|z_{\rm p}) \int dz' \phi(z'|z_{\rm p}') \int \frac{d k k^2}{(2 \pi)^3 }  \nn \\
 & \times \sum_{l,m} 4 \pi i^l j_l(ks) Y_{lm}(\hat{\bm{s}} ) \int d \hat{\bm{k}}   Y^*_{lm}(\hat{\bm{k}})  P(\bm{k}, z,z') .
\end{align}
Note that we have only one integral involving the spherical Bessel function because we expand the plane wave for $\bm{s} $.


 Eq.~\eqref{eq:w_photoz}  accounts for the photo-$z$ uncertainty exactly and it is valid for the curved sky. We only need to plug in the appropriate  power spectrum  $ P(\bm{k} , z,z')$.  However, the redshift-space power spectrum in the distant observer limit is often used, and so this effectively limits the results to the flat sky case. As long as the local line of sight (LOS) is used, the effect is mild, e.g.~\cite{YooSelak2015}.

In general, because of the azimuthal symmetry about the LOS direction $\hat{\bm{e}}$,  the power spectrum can be expanded about the LOS direction as
\begin{align}
  \label{eq:Pk_LegendreExpansion}
  P(\bm{k}, z,z')  = \sum_{{\rm even }\ell} P_\ell(k, z,z' ) \mathcal{L}_\ell( \hat{\bm{k}} \cdot  \hat{\bm{e}} ),  
\end{align}
where  $ \mathcal{L}_\ell $ is the Legendre polynomial and  $P_\ell $ is the multipole coefficient. Parity invariance further restricts the sum to even $\ell$ only.  Then the angular correlation can be written as
\begin{align}
  w( \theta, z_{\rm p},z_{\rm p}' ) & = \sum_\ell i^\ell   \int dz \phi(z|z_{\rm p})   \int dz' \phi(z'|z_{\rm p}')  \mathcal{L}_\ell( \hat{\bm{s}} \cdot \hat{\bm{e}} )  \nn \\
 & \quad \quad  \times \int \frac{d k k^2  }{ 2 \pi^2  } j_\ell(ks) P_\ell(k,z,z') ,    
\end{align}
where $j_l$ is the spherical Bessel function. To speed up the computation, the $j_\ell$ integrals are evaluated using {\tt FFTLog  } \citep{FFTLog}.

For  the linear redshift-space power spectrum with linear bias \citep{Kaiser87}
\begin{align}
  P( \bm{k}, z, z' ) & = [ b + f  ( \hat{\bm{k}} \cdot  \hat{\bm{e}} )^2 ]  [ b' + f'  ( \hat{\bm{k}} \cdot  \hat{\bm{e}} )^2 ]  \nn \\
 &  \times D(z) D(z') P_{\rm m}(k,0,0) ,  
\end{align}
where $b$ is the linear bias,  $f$ is $d \ln D /d \ln a $ with $ D $ being the linear growth factor, and  $P_{\rm m}(k,0,0) $ is the matter power spectrum evaluated at $z=0$, we have \citep{Hamilton1992,ColeFisherWeinberg1994,CrocceCabreGazta_2011}
\begin{align}
  \label{eq:w_Kaiser_linbias} 
  w( \theta,& z_{\rm p},z_{\rm p}' )  =  \int dz \phi(z|z_{\rm p})   D(z)    \int dz' \phi(z'|z_{\rm p}')  D(z')     \nn \\
 &  \times  \sum_{\ell =0,2,4} i^\ell     A_\ell   \mathcal{L}_\ell( \hat{\bm{s}} \cdot \hat{\bm{e}} )   \int \frac{d k k^2  }{ 2 \pi^2  } j_\ell(ks) P_{\rm m} (k,0,0),     
\end{align}
with
\begin{equation}
  \label{eq:Al_Kaiser}
A_\ell(z_{\rm p},z_{\rm p}') = \left\{
\begin{array}{cl}
 bb' + \frac{1}{3}( b f' + b'f ) + \frac{1}{5} f f'   & \text{for } \ell =0,\\
\frac{2}{3} (bf' + b'f )  + \frac{4}{7} ff'   & \text{for } \ell = 2,\\
\frac{8  }{35  } f f'  & \text{for } \ell=4.
\end{array} \right.
\end{equation}

In DES Y3, an anisotropic BAO damping factor is applied to the BAO feature, i.e.
\begin{align}
  \label{eq:Pk_anisotropic_damp}  
  P( \bm{k}, z, z' ) & = [ b + f  ( \hat{\bm{k}} \cdot  \hat{\bm{e}} )^2 ]  [ b' + f'  ( \hat{\bm{k}} \cdot  \hat{\bm{e}} )^2 ]  D(z) D(z')  \nn \\
 &  \times [ ( P_{\rm lin} - P_{\rm nw} )e^{ -k^2 \Sigma^2(\mu) }  + P_{\rm nw} ],  
\end{align}
where $ P_{\rm lin}$ and $ P_{\rm nw} $ denotes the linear power spectrum with and without BAO wiggles respectively. Because $\Sigma $ is a function of $\mu$, we can no longer express the first three multipoles analytically, instead we compute them numerically.  As our primary goal here is to model the BAO feature,  we shall use Eq.~\eqref{eq:Pk_anisotropic_damp}. But we will use  Eqs.~\eqref{eq:w_Kaiser_linbias}  and \eqref{eq:Al_Kaiser}  to measure the bias parameters. 

\begin{figure}
\centering
\includegraphics[width=\linewidth]{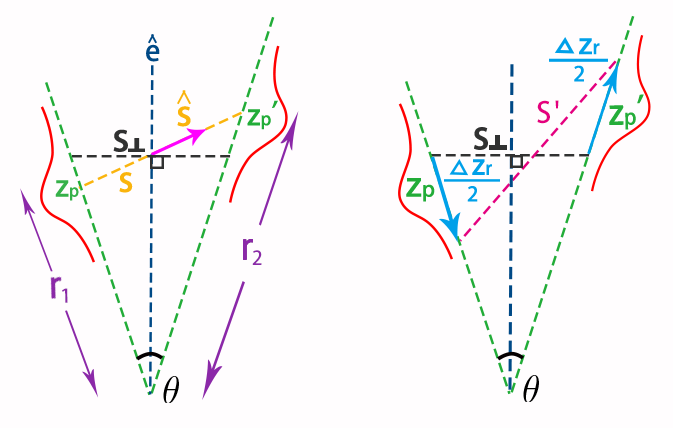}
\caption{  Left panel: The relation between the angular correlation function specified by two true redshift distributions conditional on photo-$z$ $z_{\rm p}$ and $z_{\rm p}' $, subtending an angle $ \theta$ at the observer.  The corresponding 3D correlation  function is described by the separation vector $ \bm{s} $ relative to the line of sight direction $\hat{\bm{e}} $. Right panel: The integral over the two distributions can be approximated by a 1D integral over $\Delta z_{\rm r } $ or the separation $s'$.    }
  \label{fig:w_photoz_cartoon}
\end{figure}

\subsection{ From  $w$ to $\xi_{\rm p}$ }
\label{sec:wp_to_xip}
Because of the statistical rotational invariance, the angular correlation function $w(\theta, z_{\rm eff} ) $ for the sample at the effective redshift $z_{\rm eff}$  can be parametrized by the angle $\theta$ subtended by two points on the sky at the observer.  If in addition, the redshifts (or photo-$z$) of these two points are also specified, e.g.~in the form of tomographic bins,  $w (\theta, z, z' ) $ furnishes a representation of the 3D correlation function.  On the other hand, the 3D correlation function  $\xi$ in real space depends only on the distance between the two points,  $s =  |\bm{r}_1 -   \bm{r}_2| $, thanks to the statistical translational and rotational invariance.  The redshift space distortion along the LOS direction breaks the rotational invariance  in redshift space and  $\xi$ is  usually parametrized as a function of  $ s$  and $ \mu$, the latter is  the dot product between direction of the separation, $\hat{\bm{s}}$ and the LOS $\hat{ \bm{e}}$, or in terms of  $ s_{\perp}$ and $ s_{\parallel} $, the component of $\bm{s}$ perpendicular and parallel to the LOS.  In practice, both $w$  and $\xi$ can be computed by assigning density to a grid (2D pixelized grid cells for $w$ and 3D grid cells for $\xi$) or by comparing pair counts between the data catalog and the random one.

Because $\delta_{(2)} = \delta_{(3)}$, it is clear that the general cross angular correlation function and the 3D correlation are equal provided a cosmology is assumed to convert angle and redshift to distance \change{ and the redshift bin is fine enough so that the projection effect along the LOS direction is negligible }.  Depending on how it is parametrized, the correlation function goes by different names.  We shall consider the 3D correlation function  $ \xi_{\rm p} ( s, \mu )  =  \langle  \delta_{\rm p} ( \bm{r}_1 ) \delta_{\rm p} ( \bm{r}_2 )  \rangle  $ and its projective version characterized by the transverse scale,  $\xi_{\rm p} ( s_\perp, \mu ) \equiv \langle  \delta_{\rm p} ( \bm{r}_1 ) \delta_{\rm p} ( \bm{r}_2 )  \rangle_\parallel  $, where the notation $\langle \dots \rangle_\parallel $ means that we average over $s_\parallel \equiv s \mu$ by stacking together the pairs with the same $ s_\perp \equiv  s \sqrt{ 1 - \mu^2 } $.

We obtain $\xi_{\rm p}(s,\mu)$ by mapping from $ w $.  We first evaluate the general cross correlation between redshift bin $i$ and $j$ separated by angle $\theta_k$ bin, $w_{ij}(\theta_k)$.  From its redshift bins and angular separation,  $ s$  and $ \mu$ for this pair can be obtained.  This allows us to allocate this angular correlation into the  appropriate  $s$ and $\mu $ bins. The relation between $w$ and $\xi_{\rm p } $ is schematically shown on the left panel of Fig.~\ref{fig:w_photoz_cartoon}.   Because $\xi_{\rm p}$ is measured by averaging over all possible grid cell pairs (or data pairs in the language of pair counting) in the survey data region, we loop over all the $w_{ij}(\theta_k)$ pairs that fit into the data region and give rise to the same $s$ and $\mu $ bins.  Explicitly, we have 
\begin{align}
  \label{eq:xi_w_weighting} 
\xi_{\rm p} (s,\mu) = \frac{\sum_{ ijk } f_{ ijk } w_{ij} ( \theta_k ;  z_i, z_j  )   }{\sum_{ ijk } f_{ ijk } } , 
\end{align}
where $f_{ijk}$ denotes the weight for all the cross bin pairs $ w_{ij}(\theta_k)$  falling into the $s$  and $\mu $ bins.  To approximately mimick the pair counting in the computation of $\xi_{\rm p} $, we shall use the weight $f_{ijk} = N_i N_j$  with $N_i$ ($N_j$) being the number of particles in bin $i$ ($j$).  More accurate weighting can be achieved by counting pairs as in the case of random catalog, but it would be much more computationally expensive.

Here we describe binning into $s$ and $\mu$ bin as an example. For $\xi_{\rm p}(s_\perp, \mu)$, the procedures are the same except that we compute $s_{\perp } $ from the redshifts of the bin pairs and their angular separation, and then bin the correlation based on  $s_{\perp } $ and $\mu$.

The measurement of the $\xi_{\rm p}$ statistics requires a fiducial cosmology input, while the tomographic bin correlation in term of angle and redshift can evade the cosmology dependence \citep{Montanari:2012me}.  However, the advantage of the  $\xi_{\rm p}$ statistic is that it can effectively summarize the information in multiple tomographic bins in a single correlation measurement, and it is easier to borrow the techniques developed for the spectroscopic data analysis.



We turn to comment on the definition of the LOS direction associated with a pair of galaxies on the sky.  For the angular correlation function, a natural definition of the LOS direction is the angle bisector of the angle subtended by the pairs at the observer.    In this ``angle-centric'' definition, the dot product between LOS direction $\hat{\bm{e}}_\theta $ and the unit separation vector $\hat{\bm{s}} $ can be expressed as
\beq
\hat{\bm{e}}_\theta \cdot \hat{\bm{s}}  = \frac{r_2 - r_1}{ 2 s } \sqrt{\frac{ (r_1 + r_2 + s) (r_1 + r_2 - s)  }{ r_1 r_2  }  }, 
\eeq
where $r_1$ and $r_2$ are the distances to the galaxy pair respectively, and $s$ is the length of the separation vector. On the other hand, it is natural to align the LOS with the mid-point of the separation for the 3D corrrelation function. For this  ``separation-centric'' viewpoint, the dot product between LOS direction  $\hat{\bm{e}}_s $ and $\hat{\bm{s}} $ can be written as
\beq
\hat{\bm{e}}_s \cdot \hat{\bm{s}}  = \frac{ (r_2 - r_1) (r_2 + r_1)  }{ s \sqrt{2(r_1^2 + r_2^2 ) - s^2 }}. 
\eeq
We have checked that these two defintions are essentially identical for our purpose. For example, at $z_{\rm eff} = 0.8$ and $s=100 \MpcOh$, the agreement between these two defintions are better than $10^{-3}$. Thus we can use them interchangeably.

\section{ Understanding the impact of photo-$z$ by a Gaussian case study }

\label{sec:Gaussian_photoz_limit}

\cite{Ross:2017emc} used a simple Gaussian setup to study the impact of photo-$z$ uncertainty on the BAO information. The apparatus can be constructed by the following arguments.  Suppose that the position vectors of two points are $ \bm{r}_1 $ and $ \bm{r}_2 $ and due to photo-$z$ error, they have additional displacement $\Delta \bm{r}_1 $ and $\Delta \bm{r}_2 $. The total separation vector is
\begin{align}
  \bm{s}_{\rm p}  & = ( \bm{r}_2 -  \bm{r}_1 )  +  ( \Delta \bm{r}_2 - \Delta \bm{r}_1) \nn \\
     & = \bm{s}_{\rm t \parallel}  +  \Delta \bm{r}_2 - \Delta \bm{r}_1 +  \bm{s}_{\rm t \perp} ,
\end{align}
where $ \bm{s}_{\rm t \perp} $ and $ \bm{s}_{\rm t \parallel} $  denote the perpendicular and parallel components of  $  \bm{r}_2 -  \bm{r}_1 $ w.r.t.~the LOS direction without photo-$z$ error. The photo-$z$ error only affects the direction parallel to the LOS in the plane-parallel limit. If we assume that  $ \Delta \bm{r}_1 $  and $ \Delta \bm{r}_2 $ are independent Gaussian random variables with zero mean and constant variance $\sigma^2$, then  $ \Delta \bm{r}_2 - \Delta \bm{r}_1 $ is Gaussian distributed with zero mean and variance $2 \sigma^2$. Thus we can model the effects of the Gaussian photo-$z$ error on the 3D correlation function by a 1D Gaussian distribution.

It is instructive to check under what conditions our general expression would reduce to this result.  We will see that this exercise gives valuable insight into our results.

We now assume that  the conditional true redshift distributions in Eq.~\eqref{eq:xiphoto_def} are Gaussian, namely
\beq
\phi(z| z_{\rm p}) = P_{\rm G} ( z- z_{\rm p} ; 0, \sigma^2 ) \equiv  \frac{1 }{ \sqrt{2 \pi \sigma^2} } e^{ - \frac{ (z - z_{\rm p})^2 }{2 \sigma^2 }  }. 
\eeq
By adopting the variables
\beq
z  = z_{\rm p} + \Delta z, \quad   z' =  z_{\rm p}' + \Delta z', 
\eeq
$w_{\rm p}$ can be written as
\begin{align}
  w_{ \rm p}( \theta, z_{\rm p} ,z_{\rm p}' )
  &=  \frac{ 1}{ 2 \pi \sigma^2 }     \int d \Delta z \, e^{- \frac{ \Delta z^2  }{2 \sigma^2  }  }  \int d \Delta z' \, e^{- \frac{ \Delta z'^2  }{2 \sigma^2  }  } \nn \\
 &  \times w(\theta, z_{\rm p} + \Delta z,  z_{\rm p}' + \Delta z' ).
\end{align}
After further changing to  the ``Center-of-Mass'' variables 
\begin{align}
\Delta  z_{\rm c} =  \frac{\Delta z + \Delta z'}{2} , \quad  \Delta z_{\rm r} = \Delta z  -  \Delta z',
\end{align}
we have 
\begin{align}
  w_{\rm p}( \theta, z_{\rm p}, z_{\rm p}' ) & = \frac{1}{ 2 \pi \sigma^2 }   \int d \Delta z_{\rm c}  \int d \Delta z_{\rm r}         \exp \bigg( - \frac{ \Delta z_{\rm c}^2  }{ \sigma^2 }  -   \frac{ \Delta z_{\rm r}^2  }{ 4\sigma^2}   \bigg)   \nn  \\
 & \quad  \times w \Big( \theta, z_{\rm p} + \Delta z_{\rm c} + \frac{ \Delta z_{\rm r} }{ 2 },   z_{\rm p}' + \Delta z_{\rm c} - \frac{ \Delta z_{\rm r} }{ 2 }   \Big).
\end{align}
To proceed, we neglect the $ \Delta z_{\rm c} $ dependence in $w$ as it gives an overall shift of the pair in redshift \footnote{The standard 3D correlation analysis neglects the redshift evolution within the redshift bin. } and we arrive at 
\begin{align}
  \label{eq:wp_Gaussian_approximation}
  w_p( \theta, z_{\rm p}, z_{\rm p}' ) 
  &  \approx     \int d \Delta z_{\rm r}   \frac{1}{\sqrt{ 4 \pi\sigma^2 }}   \exp \bigg( -   \frac{\Delta z_{\rm r}^2 }{ 4  \sigma^2 }   \bigg)  \nn \\
 &  \times w \Big( \theta,  z_{\rm p} + \frac{ \Delta z_{\rm r} }{ 2 },   z_{\rm p}' - \frac{ \Delta z_{\rm r} }{ 2 }  \Big) . 
\end{align}
By mapping to $s$ and $\mu$,  we can get the corresponding equation for $\xi_{\rm p}  $, i.e.
\begin{align}
  \label{eq:xip_Gaussian_approximation}
    & \xi_{\rm p}\big( s(\theta, z_{\rm p}, z_{\rm p}'), \mu(\theta, z_{\rm p}, z_{\rm p}')  \big) 
   \approx    \int d \Delta z_{\rm r}  \, P_{\rm G}(\Delta z_{\rm r}; 0, 2 \sigma^2)  \nn \\
& \times  \xi \Big( s' \big(\theta, z_{\rm p} + \frac{ \Delta z_{\rm r} }{ 2 },   z_{\rm p}' - \frac{ \Delta z_{\rm r} }{ 2 } \big), \mu' \big(\theta, z_{\rm p} + \frac{ \Delta z_{\rm r} }{ 2 },   z_{\rm p}' - \frac{ \Delta z_{\rm r} }{ 2 } \big)  \Big) . 
\end{align}
This equation invites the interpretation on the right panel of Fig.~\ref{fig:w_photoz_cartoon}.

Alternatively we can use the separation distance $s'$ (shown on the right panel of Fig.~\ref{fig:w_photoz_cartoon}) as the integration variable
\begin{align}
  \label{eq:xip_Gaussian_approximation_sp}
   & \xi_{\rm p}\big( s(\theta, z_{\rm p}, z_{\rm p}'), \mu(\theta, z_{\rm p}, z_{\rm p}')  \big)  \nn \\
   \approx  &   \int ds' \left( \frac{ds'}{ d \Delta z_{\rm r} } \right)^{-1}   \, P_{\rm G}(\Delta z_{\rm r}(s'); 0, 2 \sigma^2)   \xi \big( s', \mu'  \big).
\end{align}
We can think of $s'$ as a function of $\Delta z_{\rm r} $. When $\Delta z_{\rm r} $ increases from the negative end,  $s'$ first decreases and it reaches the minimum value  $ s' = s_\perp $ before starting to increase.  The scale $ s' = s_\perp $ divides the integral into two parts, and they should be evaluated separately.

The inverse of the Jacobian is given by
\beq
\label{eq:Jinv_dsp_dDzr}
\frac{d s'  }{ d \Delta z_{\rm r} } = \frac{c }{2 s' } \left[ \frac{ r(z_2) }{ H(z_2)} - \frac{ r(z_1) }{ H(z_1)} - \cos \theta \left(  \frac{r(z_1)  }{ H(z_2) } - \frac{ r(z_2) }{H(z_1)}   \right) \right], 
\eeq
where  $r(z)$ and $H(z)$ respectively represent the comoving distance and Hubble rate evaluated at redshift $z$, $c$ is the light speed, and $ z_1$ ($z_2$) denotes  $ z_{\rm p} - \frac{ \Delta z_{\rm r} }{ 2 } $ ($z_{\rm p}' + \frac{ \Delta z_{\rm r} }{ 2 }$). It is important to note that this derivative vanishes when $ z_1 = z_2 $, which corresponds to $s' = s_\perp$.

\begin{figure*}
\centering
\includegraphics[width=\linewidth]{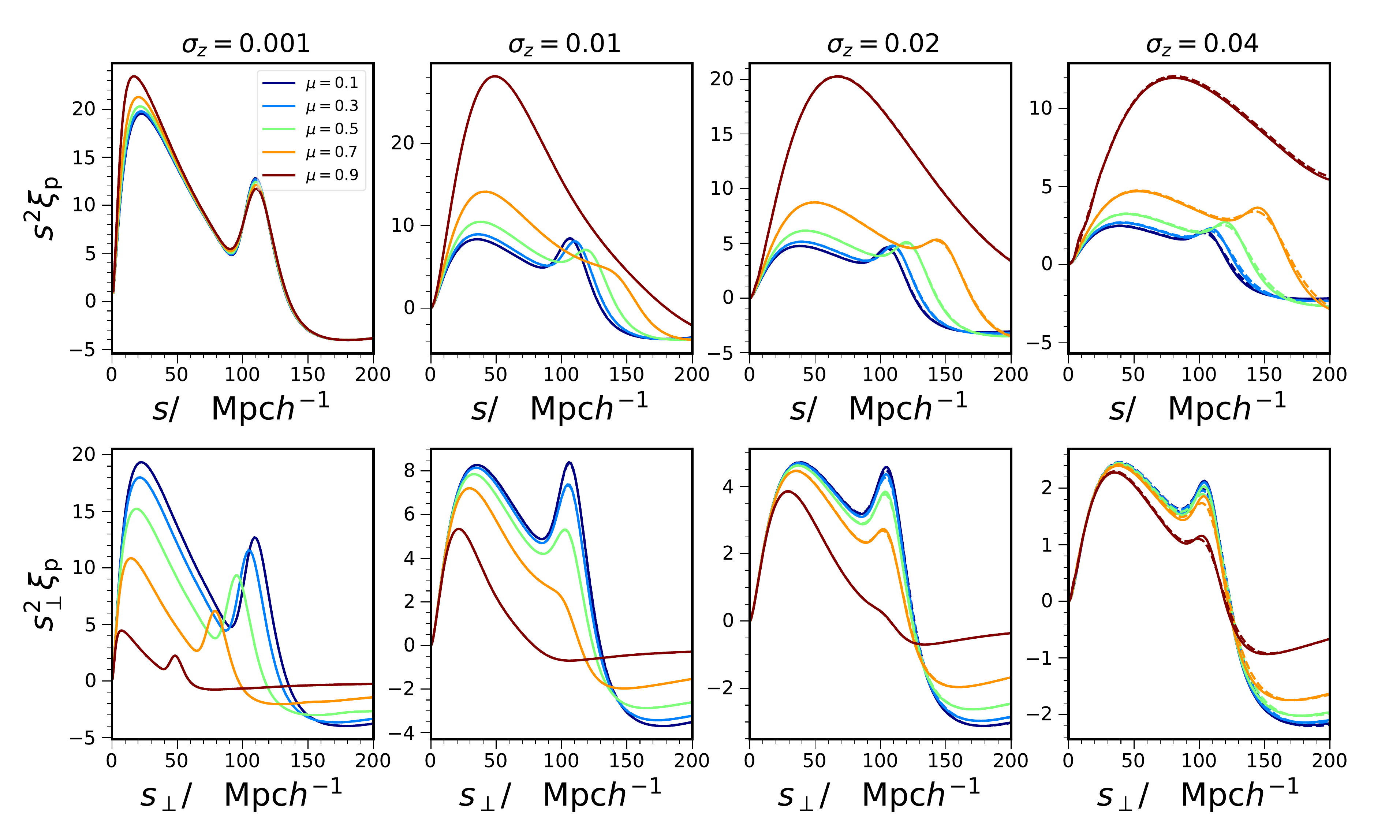}
\caption{ The wedge correlation function computed with Gaussian photo-$z$ distributions.  The upper and low panels show the results for using the separation $s$ and the transverse scale $s_\perp$ as the independent variables respectively.  The 1D approximation Eq.~\eqref{eq:xip_Gaussian_approximation_sp} (solid) and the exact 2D integral results (dashed) agree with each other well, but small deviation occurs for large value of $\sigma_z$.  Four values of $\sigma_z$ are displayed (left to right panels). Different colors indicate wedge correlation of different values of $\mu$. }
\label{fig:xip_wedge_full}
\end{figure*}

Eq.~\eqref{eq:xip_Gaussian_approximation} reveals that the photo-$z$ correlation function is a projection of the underlying true correlation with a Gaussian weight of zero mean and variance $2  \sigma^2$, in agreement with Eq.~(6) of \cite{Ross:2017emc}.  In Fig.~\ref{fig:xip_wedge_full}, we show the wedge correlation function obtained with the Gaussian photo-$z$ distribution.  The effective redshift of the two conditional true redshift distributions centered at $z_{\rm p}$ and $z_{\rm p}'$ respectively, is taken to be $z_{\rm eff} = 0.8$. The standard deviation of the distribution is assumed to be $ \sigma =  ( 1+ z_{\rm eff} ) \sigma_z $ and a suite of $ \sigma_z$ values are considered: 0.001, 0.01, 0.02, and 0.04. In this plot, for simplicity the linear real-space correlation is used.  We have compared the results computed with the approximation Eq.~\eqref{eq:xip_Gaussian_approximation_sp}  and  by directly integrating over the 2D integral. We find that the 1D approximation is in good agreement with the exact result, but deviations become apparent for high $\sigma_z $.

\begin{figure*}
\centering
\includegraphics[width=\linewidth]{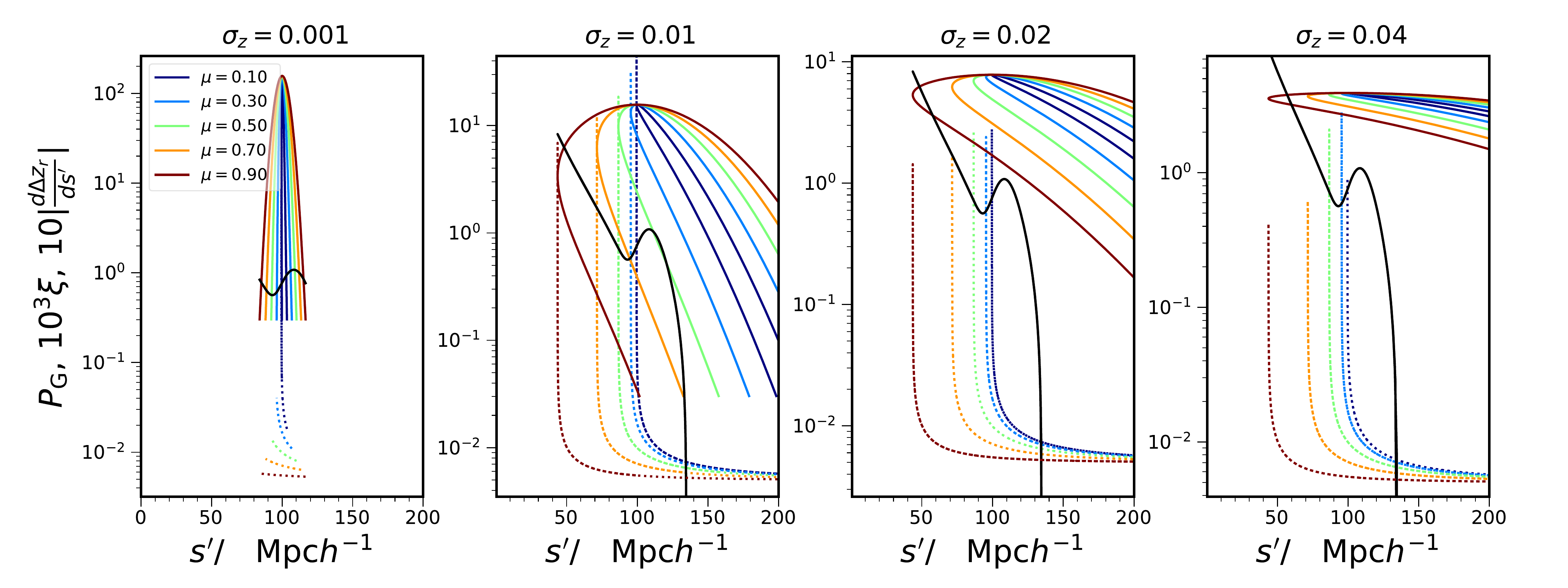}
\caption{ The factors in the photo-$z$ projection integral [Eq.~\eqref{eq:xip_Gaussian_approximation_sp}] as a function of the separation $s'$.  As shown on the right panel of Fig.~\ref{fig:w_photoz_cartoon}, as $\Delta z_r $ increases, $s'$ first decreases until it reaches the minimum value $s' = s_\perp $ and then it increases.  Both the Gaussian weight $ P_{\rm G} $ (solid, color) and absolute value of the Jacobian $ | d \Delta z_{\rm r}/ ds'|$ (multiplied by a factor of 10, dotted, color) correspond to the paths traversed as $ \Delta z_{\rm r} $ increases. The point where all the $ P_{\rm G }$ curves with different $\mu$ meet corresponds to $ s' = s( z_{\rm p}, z_{\rm p}')$ (chosen to be  $100 \MpcOh $ here).  The Jacobian diverges at $ s' = s_\perp $ and this indicates as the vertical line in the plot. The verticle part for the Jacobian touches the $P_{\rm G} $ curve  as the latter turns at  $s' = s_\perp $.  The underlying correlation function (multiplied by a factor $10^3$, solid black) is also overplotted. See the text for further explanations. }
  \label{fig:xip_intg_components}
\end{figure*}

Interestingly, while for small $\sigma_z$ such as 0.001, the BAO bump peaks at the true sound horizon size, $s_{\rm sh}$, it shifts to larger values as $\sigma_z$ increases. Also for large $\sigma_z $, it is clear that the BAO bump moves to larger $s$ as $\mu$ increases.  In the lower panels of Fig.~\ref{fig:xip_wedge_full}, we show  $\xi_{\rm p} $ as a function of the transverse scale $s_\perp  = s \sqrt{ 1- \mu^2 }  $. As observed in \cite{Ross:2017emc}, when  $ \xi_{\rm p} $ is plotted against $ s_\perp  $, the BAO position appears to line up at  $ s_\perp = s_{\rm sh} $ for $ \sigma_z \gtrsim 0.02 $. Thus this plot suggests that the BAO peak in the upper panel (the usual way to plot $\xi$) is directly related to the scale $ s_\perp $ for $ \sigma_z \gtrsim 0.02 $.


To shed light on these results, we show in Fig.~\ref{fig:xip_intg_components} the three factors of the integrand of Eq.~\eqref{eq:xip_Gaussian_approximation_sp}: the absolute value of the Jacobian $ | d \Delta z_{\rm r}  / d s' |   $, the Gaussian weight $ P_{\rm G} $  and the underlying correlation function $\xi$. We shall shortly explain these terms in details.  In this illustration, we take $ s(z_{\rm p},z_{\rm p}') = 100 \MpcOh $.

For small $\sigma_z$ such as $ 0.001 $, which is close to the spectroscopic redshift case, the Gaussian weight is narrowly distributed about the scale $s(z_{\rm p},z_{\rm p}')$. The distribution $P_{\rm G}$ as a function of $s'$ is in fact mildly tilted for small $\mu$ and it becomes more symmetric when $\mu $ increases.  As we mentioned, Eq.~\eqref{eq:Jinv_dsp_dDzr} vanishes at $s'=s_\perp $, and hence the Jacobian  diverges. Except for small $\mu$ such as $ \mu \lesssim 0.1 $, the divergence occurs at the scale where $P_{\rm G}$  is essentially zero and hence it is irrelevant.  Therefore, in this case the BAO scale in $\xi_{\rm p}$ is controlled by $P_{\rm G}$, and it corresponds to the true sound horizon well.  

As $\sigma_z$ increases, the $P_{\rm G} $ distribution becomes more tilted.  Among the cases shown, for $ \sigma_z \ge 0.01 $, $s'$ is clearly not a monotonic function of $\Delta z_{\rm r} $, as evident from the  multi-valued $ P_{\rm G} $ curves.  The scale where all the $\mu$-curves pass through corresponds to $s' = s$. The smallest $s'$ that each $\mu$-curve attains corresponds to the transverse scale $s_\perp = s\sqrt{ 1 - \mu^2}$.   The Jacobian formally diverges at $s' = s_\perp $. For small $\mu $,  as $s_\perp$ coincides with $s$ well,  the Jacobian and the Gaussian weight both peak at $s'=s$, the value of the integral is essentially equal to  $\xi(s)$.

For larger $\mu$, it depends on the value of $\sigma_z$. While the Jacobian is sharply peaked at $s'=s_\perp$ and $\xi$ is also rapidly decreasing with $s'$.  For $\sigma_z \lesssim 0.01 $, the Gaussian weight at $s'=s_\perp$  is suppressed relative to its maximum value at $s$.  Thus the integral is not well peaked at $ s' = s_\perp $. The distribution is more tilted as  $\sigma_z$ further increases and  becomes almost horizontally laid down for  $ \sigma_z \gtrsim 0.04 $. Because the Gaussian weight suppression is much more mild for  $\sigma_z \gtrsim 0.02 $, the integral is essentially given by  $ \xi(s_\perp) $. This explains why the scale $s_\perp$ but not $s$, is a good indicator of the underlying correlation function when  $ \sigma_z \gtrsim 0.02 $.


Although we have carried out the calculations explicitly using Gaussian conditional true redshift distribution, the lessons learned are more general. The approximation that the $z_{\rm c}$ dependence can be integrated out and the final results can be approximated by a 1D projection integral should be valid for well-behaved unimodal distributions (modulo small random fluctuations). In particular, the fact that the underlying correlation is reflected by $ s_\perp $ primarily stems from the divergence of the Jacobian, which is not dependent on the precise form of the distribution.

We mention that the divergent Jacobian plays a similar role as the Limber approximation \citep{Limber1953,LoVerde:2008re}, which is phrased as a linear relation between the spherical wavenumber $\ell $ and the Fourier mode $k$ and is effective in reducing the dimension of the projection integrals.   In this approximation,  although only the  Fourier modes transverse to the line of sight are included,  it works very well for reasonably large $\ell$ because those are the only significant contributions to the integral.

As $\xi_{\rm p} $ corresponds to the underlying correlation function at the transverse scale $s_\perp $ for $ \sigma_z  \gtrsim 0.02 $, we can stack all the modes of $ \xi_{\rm p} $ with the same  $s_\perp $ together to increase the signal-to-noise. Thus in this sense, the 3D $\xi_{\rm p} $ characterized by $ s_\perp $ is effectively a projected correlation function.   Consistent with \citet{SeoEisenstein_2003,BlakeBridle_2005},  $\xi_{\rm p}(s_\perp) $ for  $\sigma_z \gtrsim 0.02 $  only measures the transverse BAO and cannot be used to probe the radial BAO and hence the Hubble parameter directly.



\section{Gaussian covariance }
\label{eq:Gaussian_covmat} 
In this section, after briefly reviewing the Gaussian covariance for the angular correlation function, we then employ it to derive the Gaussian covariance for $\xi_{\rm p}$.

In the case of complete sky coverage, $\delta_{\rm p} $ in Eq.~\eqref{eq:delta_p} can be expressed in terms of the spherical harmonics as
\beq
\delta_{\rm p}(z_{\rm p},\bm{r}) = \sum_{ \ell m} a_{\ell m}(z_{\rm p})  Y_{\ell m}( \hat{\bm{r}} ).  
\eeq
We can define an estimator for angular correlation function:
\begin{align}
\hat{w}_{ij}\big(  \cos^{-1}( \hat{\bm{r}} \cdot \hat{\bm{r}}' ) \big)  = \sum_{ \ell m }  & \sum_{\ell'm'}   Y_{\ell m}( \hat{\bm{r}} )  Y_{\ell'm'}( \hat{\bm{r}}' )  \nn \\
& \times a_{\ell m}(z_{{\rm p}_i})  a_{\ell' m'}(z_{{\rm p}_j}) .  
\end{align}
Its expectation value gives \citep{Peebles} 
\begin{align}
  w_{ij}(  \theta )  
  =    \sum_\ell \frac{ 2 \ell + 1 }{ 4 \pi } \mathcal{L}_\ell ( \cos \theta) C_\ell^{ij} , 
\end{align}
where   $ C_l^{ij} $ is the cross angular power spectrum defined as 
\beq
\label{eq:Cl_def}
\langle  a_{\ell m}(z_{{\rm p}_i} )  a_{\ell' m'}^*( z_{{\rm p}_j} ) \rangle  = \Big( C_\ell^{ij} + \frac{ \delta_{\rm K}^{ij} }{ \bar{ n }_i }  \Big) \delta^{\ell \ell'}_{\rm K}  \delta^{m m'}_{\rm K},  
\eeq
with $\delta_{\rm K}^{ab}$ being the Kronecker delta. \change{  Because galaxies are discrete tracers, we have included the shot noise contribution with  $\bar{ n }_i$ being the tomographic angular number density given by
\beq
 \bar{ n }_i =  \int_{ {\rm Bin} i} dz_{\rm p}  \bar{n}_{\rm p}(z_{\rm p} ). 
\eeq
}

Similarly, we can derive its Gaussian covariance  \citep{CrocceCabreGazta_2011}
\begin{align}
  \label{eq:cov_mat_crossz}
&\mathrm{Cov}[ \hat{w}_{ij}(\theta),  \hat{w}_{mn}(\theta')] =  \sum_{\ell} \frac{(2 \ell +1)}{(4 \pi)^2} P_\ell(\cos \theta) P_{\ell}(\cos\theta') \nonumber \\
& \times \bigg[ \Big(C^{im}_{\ell}+ \frac{\delta^{im}_{\rm K}}{\bar{n}_i}\Big)   \Big( C^{jn}_{\ell}+\frac{\delta^{jn}_{\rm K}}{\bar{n}_j}\Big) + \Big(C^{in}_{\ell}+\frac{\delta^{in}_{\rm K}}{\bar{n}_i }\Big) \Big(C^{jm}_{\ell}+\frac{\delta^{jm}_{\rm K}}{\bar{n}_j}\Big) \bigg]. 
\end{align}
Poisson shot noise is assumed in this equation. To obtain the covariance for partial sky coverage, we can apply the so-called $ f_{\rm sky}$ approximation, i.e.~the covariance scaling inversely with the fraction of sky coverage $f_{\rm sky }$ \citep{Scott_etal1994}.

A few comments are in order. We evaluate the cross harmonic power spectra $C_\ell^{ij} $ using {\tt camb sources} \citep{Lewis:2007kz}.  To account for the fact that the data vector is binned into angular bins, we need to use the bin-averaged Legendre polynomial \citep{Salazar-Albornoz:2016psd}. In addition, as the terms are computed numerically we need to evaluate some of the shot noise terms analytically to ensure numerical convergence \citep{Chan:2018gtc}.

Using Eq.~\eqref{eq:xi_w_weighting}, we can then write the covariance of $\xi_{\rm p }$ in terms of that of $w$: 
\begin{align}
   &  \quad \quad \mathrm{Cov}[ \xi_{\rm p}(s, \mu), \xi_{\rm p}(s', \mu') ]  \nn \\
  &  \equiv \langle \xi_{\rm p}(s, \mu), \xi_{\rm p}(s', \mu')  \rangle   -   \langle \xi_{\rm p}(s, \mu)  \rangle    \langle  \xi_{\rm p}(s', \mu')  \rangle   \nn \\
  & = \frac{ \sum_{ ijk } \sum_{ lmn }  f_{ijk} f_{lmn} \mathrm{Cov}( w_{ij} ( \bar{z} , \Delta z, \theta_k ) , w_{lm} ( \bar{z} , \Delta z', \theta'_n )    )     }{ \sum_{ijk } f_{ijk}  \sum_{lmn } f_{lmn}  }. 
\end{align}
This enables us to map the Gaussian covariance for $w$ to that for $\xi_{\rm p} $.  Similarly, we can get the covariance for $ \xi_{\rm p}(s_\perp, \mu)$ using the corresponding weight.

\change{
  It is worth discussing more about the shot noise term. The shot noise arises from the self-correlation of the discrete particles.  In the Poisson model, the shot noise contribution to the correlation  is given by 
\begin{align}
  &  \frac{ 1 }{ \bar{n}_{\rm p} ( z_{\rm p} )   \bar{n}_{\rm p} ( z_{\rm p}' )  }   \int dz f(z| z_{\rm p} )    \langle  n(z, \hat{\bm{r}} ) \rangle  \Ddel(z_{\rm p} -z_{\rm p}') \Ddel(\hat{\bm{r}} - \hat{\bm{r}}' )    \nn  \\
  \label{eq:shotnoise_photoz}
 = &   \frac{ 1 }{ \bar{n}_{\rm p} ( z_{\rm p} ) }  \Ddel( z_{\rm p} -  z_{\rm p}' ) \Ddel(\hat{\bm{r}} - \hat{\bm{r}}' ), 
\end{align}
in which the two-point distribution is converted to a product of a one-point distribution and the Dirac delta distribution.  Thus the shot noise contribution to the 3D correlation function depends on the  3D number density  $ \bar{n}_{\rm p} $.   Although the true redshift distribution of the neighboring bins may overlap with each other for photometric galaxies, there is no cross photo-$z$ (or cross-bin in tomographic analysis) shot noise contribution because shot noise arises from self-correlation and each galaxy is assigned one and only one photo-$z$. }

\change{
  Our method approximates the 3D correlation and the shot noise by the tomographic version. For the signal,  when the tomographic bin size is small compared to photo-$z$ uncertainty, the results are convergent. For the shot noise, we approximate the Dirac delta in redshift in Eq.~\eqref{eq:shotnoise_photoz} by a Kronecker delta and the volume density by the tomographic one.  }


\section{ Mock tests }

In this section, we first test the theory template and the Gaussian covariance for $\xi_{\rm p}$ against mock catalogs in Sec.~\ref{sec:Template_and_Covariance} and then test the BAO fit pipeline and present the fit results in details in Sec.~\ref{sec:BAO_fit}.   To do so, we use the ICE-COLA mocks \citep{ COLAmocks_Ismael}, which is a dedicated mock catalog for the DES Y3 BAO analyses.  We briefly describe it here, and refer readers to \citet{COLAmocks_Ismael} for more details.

The ICE-COLA mocks are derived from the COLA simulations which is based on the COLA method \citep{Tassev_etal_2013} and implemented by the ICE-COLA code \citep{ice-cola}. The COLA method combines the second order Lagrangian perturbation theory with the particle-mesh simulation technique to ensure that the large-scale modes remains accurate when coarse simulation time steps are used.  The simulation consists of $2048^3$ particles in a cube of side length 1536 $\MpcOh$, matching to the mass resolution of MICE Grand challenge $N$-body simulations \citep{Fosalba_etal2015,Crocce_etal2015}.   To generate the whole sky light-cone simulation at  $z \sim 1.4$, the simulation box is replicated three times in each Cartesian direction (altogether 64 copies). The replication has important consequences for the covariance measurements, and we will further comment on it later on.  Mock galaxies are assigned to the ICE-COLA halos using a hybrid Halo Occupation Distribution and Halo Abundance Matching recipe, following a similar approach in \citet{Avila:2017nyy}. The mocks are calibrated with the redshift distribution and the angular clustering amplitude of the true galaxy samples.  From the full-sky light-cone mocks, four maps with the DES Y3 footprint are extracted. Thus there are 1952 mock catalogs in total, although we will only use a subset of them.

\change{ A key challenge is to assign realistic photo-$z$'s to the mock galaxies.  To do so the photo-$z$ distribution is estimated using 8362 galaxies with known spectroscopic redshift from VIPERS survey \citep{VIPERS}. Using these galaxies, which have both the spectroscopic redshift $z_{\rm s} $ and the photo-$z$ from DES, the two-dimensional distribution $P(z_{\rm p}, z_{\rm s} )$ can be estimated.  By sampling  $P(z_{\rm p}, z_{\rm s} )$ conditioned on the true redshift $ z_{\rm s}$, the candidate mock galaxies are put to the appropriate photometric bins.  A further requirement is that the mock galaxies must match the abundance of the observed photometric galaxies.   Suppose there are $N(z_{ {\rm p} i })$ galaxies in the photometric bin $z_{ {\rm p} i } $, the most luminous $N(z_{ {\rm p} i }) P(z_{ {\rm p} i}, z_{{\rm s} j} ) $ galaxies  from the bin $( z_{ {\rm p} i}, z_{{\rm s} j} )$ is taken as the mock sample. By construction, the mock galaxies will match the observed galaxy abundance and the conditional true redshift distribution. }

As the cosmology adopted by the mock catalog is the MICE cosmology, the theory templates, i.e.~the theoretical model described in Sec.~\ref{sec:wp_to_xip}, and the Gaussian covariance are computed in this  cosmology,  although we will also consider the template computed in Planck cosmology  \citep{Planck2020}. 

\subsection{ Template and covariance}
\label{sec:Template_and_Covariance}

In this subsection, we use the ICE-COLA mocks to test the theories developed in Sec.~\ref{sec:theory_template} and Sec.~\ref{eq:Gaussian_covmat}.    The theory calculations require the conditional true redshift distribution and the bias parameters measured from mocks. Thus we shall first present the measurement of these quantities before confronting the theories with measurements.  \change{ In the Appendix~\ref{sec:specz_check}, we check the template calculation using the spectroscopic correlation function. }

The spread of the conditional true redshift distribution generally increases with redshift.  \change{  As  $\phi$  depends on the number density in true redshift and the spectroscopic data is not available outside $[0.6,1.1]$,  we cannot reliably calibrate $ \phi $ for all bins. Here we approximate  $ \phi $ by $f$. The preliminary estimate suggests that  the effect is small, but when the spectroscopic density is rapidly changing such as for the high redshift bins, it should be taken into account. We shall present more detailed analysis in future work. } In the standard tomographic analysis, the samples are divided into a number of redshift bins and their auto (and cross) correlation functions are considered. For example, in the DES Y3 BAO analyses, the samples in the photo-$z$ range $[0.6, 1.1]$ are divided into five tomographic bins with bin width $\Delta z = 0.1$. For 3D correlation function, because $\phi$  is continuously sampled by the galaxy pair counts, we have to use fine photo-$z$ bins in order to sample the $\phi$  accurately and smoothly.

\begin{figure}
\centering
\includegraphics[width=\linewidth]{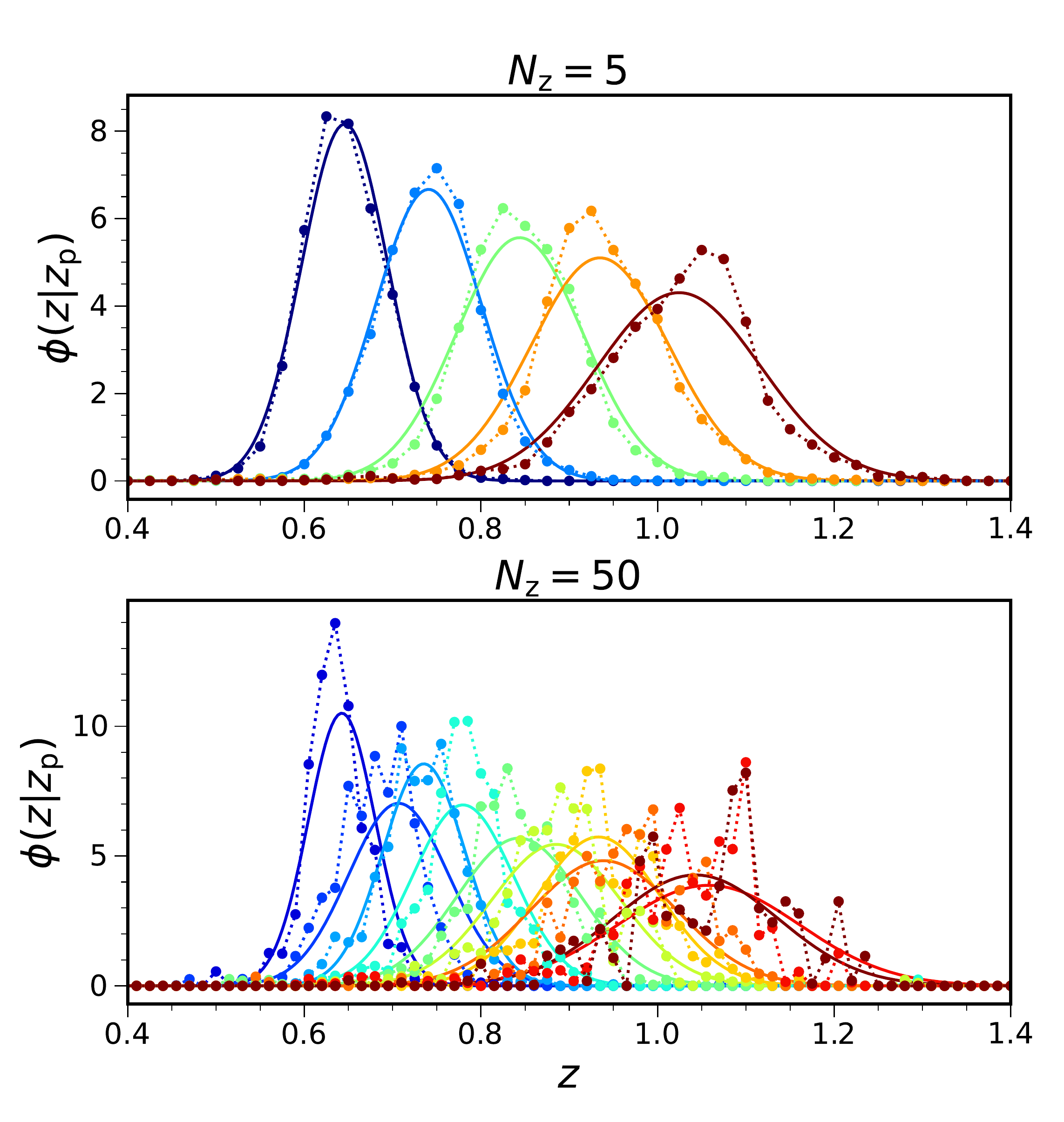}
\caption{ The conditional true redshift distributions measured from COLA mocks (filled circles)  with 5 (upper panel) and 50 (lower panel) photo-$z$ bins. The Gaussian approximation is also overplotted (solid curves). For clarity, in the case of 50  bins, only those whose bin number is divisible by 5 are shown. }
\label{fig:dndz_zbins50_5_vertical}
\end{figure}

To investigate the number of photo-$z$ bin width required, we consider a number of $\phi$'s measured from the ICE-COLA mocks in the photo-$z$ range [0.6, 1.1]. We divide this photo-$z$ range into redshift bins of equal spacing.   In Fig.~\ref{fig:dndz_zbins50_5_vertical}, we show a sample of $\phi$ obtained with 5 and 50 photo-$z$ bins. We also overplot the corresponding Gaussian approximation with the same mean and variance as the true $\phi$. The Gaussian approximation is more accurate for the five-bin case as it is averaged over a larger redshift range. At high redshift, $\phi$  is more irregular and the Gaussian approximation is less accurate.   

\begin{figure}
\centering
\includegraphics[width=\linewidth]{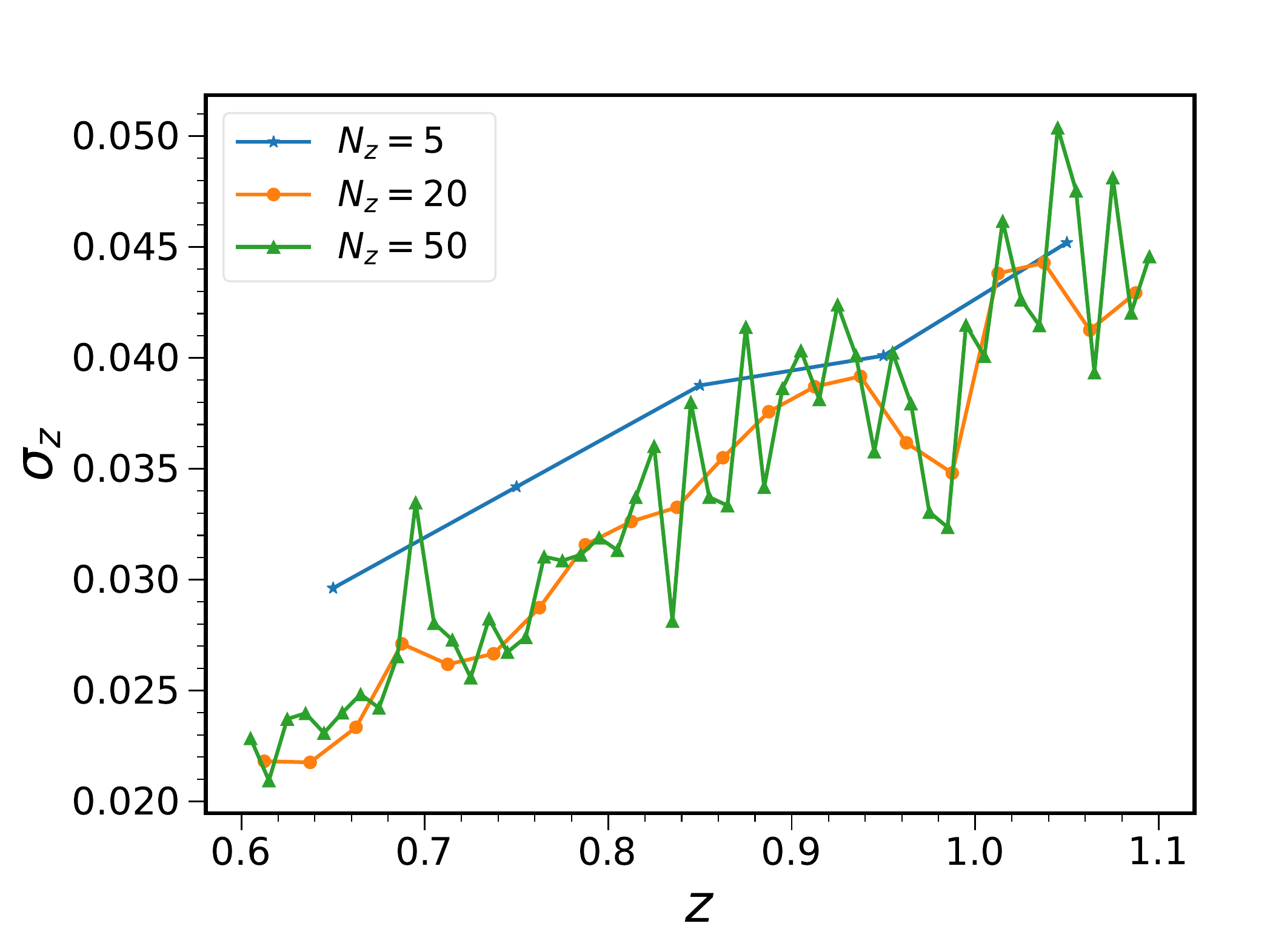}
\caption{ $\sigma_z$ [Eq.~\eqref{eq:sigma_z}] of the conditional true redshift distribution for 5 (blue), 20 (orange), and 50 (green) photo-$z$ bins.  }
\label{fig:sigmaz_Nzbins}
\end{figure}

The photo-$z$ spread is often characterized by
\beq
\label{eq:sigma_z}
\sigma_z =  \frac{\sigma_{\rm p}}{ 1+z },
\eeq
where $ \sigma_{\rm p} $  is the standard deviation of $\phi$. We measure $\sigma_z $ from  $\phi$  for different number of photo-$z$ bins and the results are displayed in Fig.~\ref{fig:sigmaz_Nzbins}.   While $ \sigma_z $ is overestimated in the 5-bin case, the 20-bin results are consistent with the ones from the 50-bin case except that the 50-bin results are more noisy.   This suggests that when the bin width $\Delta z =  0.025 \lesssim \sigma_z $, the photo-$z$ bin is fine enough to probe the intrinsic photo-$z$ uncertainty. We emphasize that $ \sigma_z$ is used for sanity checks only, we shall use the full distribution in the test of the analysis.

\begin{figure}
\centering
\includegraphics[width=\linewidth]{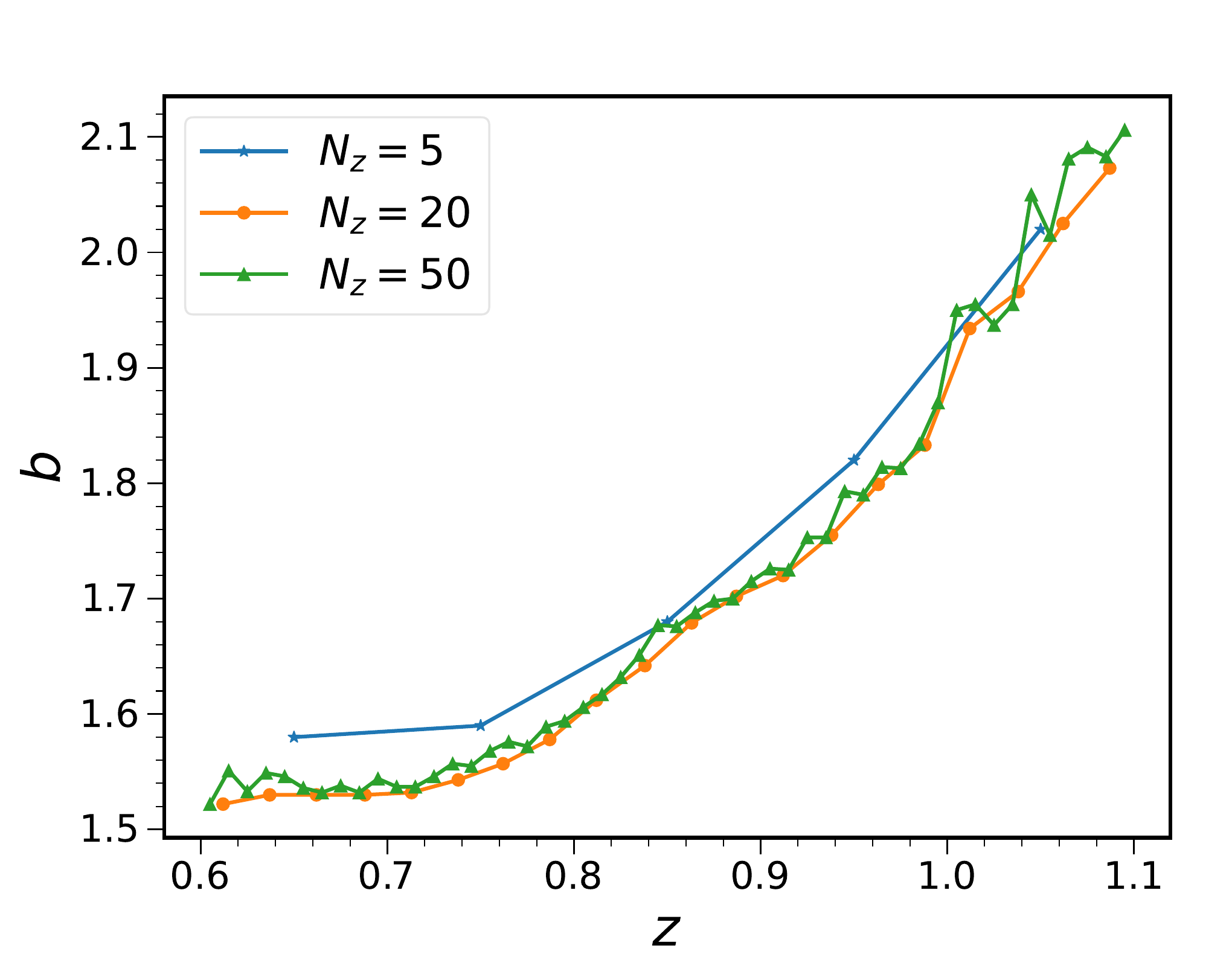}
\caption{ The bias parameters measured from 5 (blue), 20 (orange), and 50 (green) photo-$z$ bins.  }
\label{fig:bias_z_Nzbins}
\end{figure}

Another ingredient required is the bias parameters.  We measure the linear bias parameters by fitting the auto-angular correlation [Eq.~\eqref{eq:w_Kaiser_linbias} and \eqref{eq:Al_Kaiser}] to the mock measurements as in \citet{Chan:2018gtc,Abbott:2017wcz}. The fit is performed by fitting to the mean of mocks in the angular range of  [0.5,1.5] degree.  We show the bias parameters obtained with different number of tomographic bins in Fig.~\ref{fig:bias_z_Nzbins}.  Similar to the case of $ \sigma_z $, the 20-bin and 50-bin cases show consistent result with the 50-bin ones showing more random fluctuations,   while the five-bin results are slightly higher. 

\begin{figure}
\centering
\includegraphics[width=\linewidth]{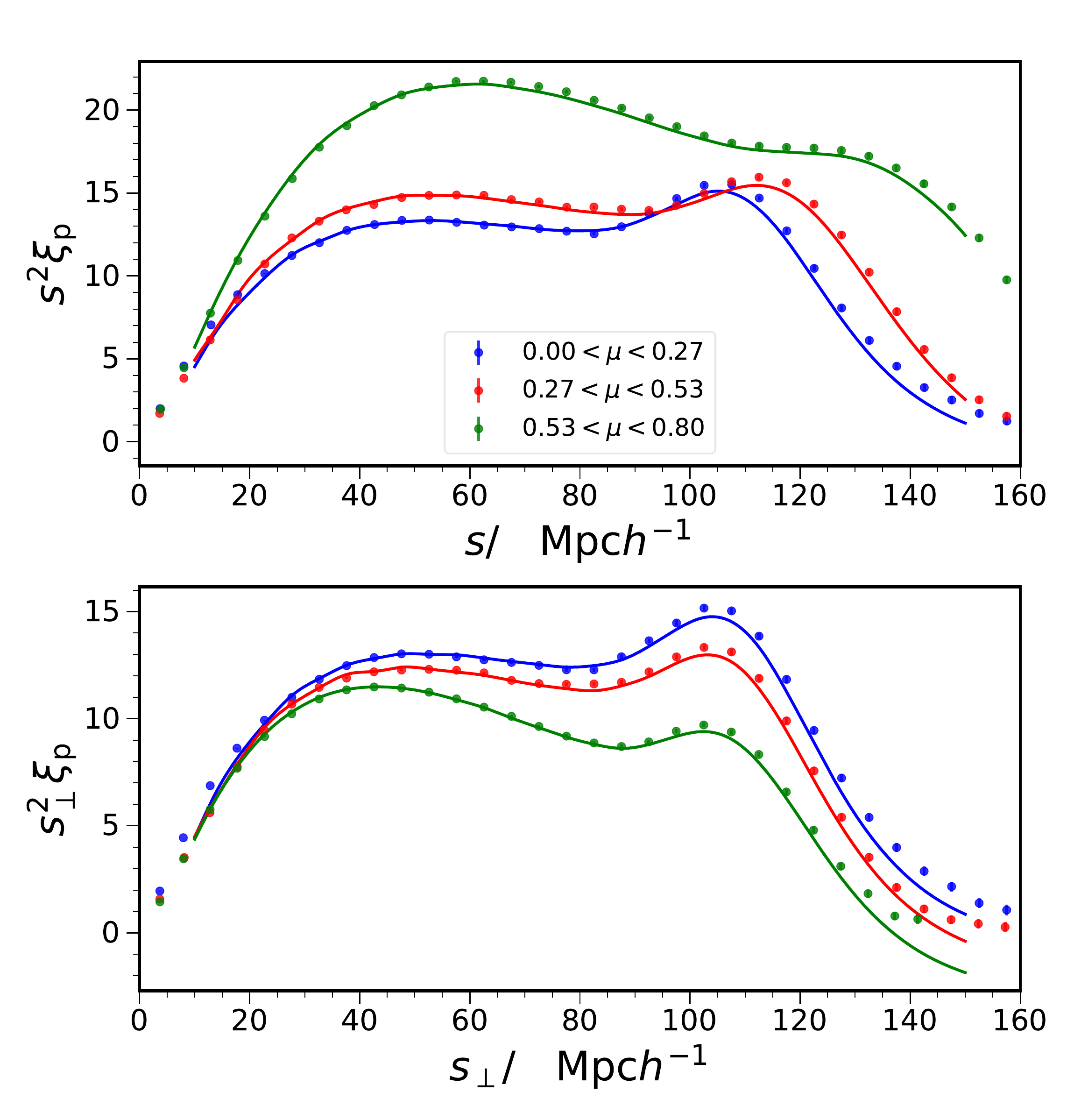}
\caption{ The wedge correlation function $\xi_{\rm p} $ as a function of the separation $s$ (upper panel) and the transverse separation $s_\perp$ (lower panel). The $\mu$ range in [0, 0.8] is divided into 3 equal spacing bin, [0,0.27] (blue), [0.27,0.53] (red), [0.53,0.8] (green).  The measurements (filled circles) are the mock mean and the error bars estimated by the standard error are too small to be seen. The theory templates are the solid curves.   }
\label{fig:xi_mock_template_B}
\end{figure}

We compare the theory template against the mock measurements in Fig.~\ref{fig:xi_mock_template_B}.   The 3D correlation $\xi_{\rm p}$ is measured  using the Landy-Szalay estimator \citep{LandySzalay_1993} via the public code {\tt CUTE} \citep{Alonso_CUTE}.  Limited by the computing power, we only measure the correlation function from 384 ICE-COLA mocks. Judging from the size of the error bars, which are estimated by the standard error of the mean, the fidelity of the results are well sufficient for our purpose here. We estimate the correlation by considering the pairs with maximum radial separation $120 \MpcOh $ and transverse separation $175 \MpcOh$.  To ensure good sampling of the projected correlation within the maximum radial separation, we only consider $\xi_{\rm p} $ up to $\mu=0.8$.   We further divide the $\mu$ range $[0, 0.8]$ into three equal spacing ``wedges''. Both $\xi_{\rm p} $ versus $s$ and $s_\perp$ are shown.

 For the theory template, we first compute $w_{ij}(\theta ) $ in 50 photo-$z$ bins with angular bin width of 0.02 degree. Because the bin width in the radial direction is wide ($\Delta r \sim 20 \MpcOh $), we  linearly interpolate the resultant angular correlation to a larger redshift grid of 500 bins ($\Delta z = 0.005$) before binning  all the pairs  $w_{ij}(\theta_k) $ into the appropriate $\xi(s,\mu)$ bin.  It is important, especially for the projected correlation function, to ensure that the maximum ranges in the radial and transverse directions used in template construction match those in the measurements.  We find that the template obtained is of slightly lower amplitude than the mock measurement (by about 3\%), the amplitude deficit is probably due to sparse sampling in the radial direction. The template shown in Fig.~\ref{fig:xi_mock_template_B} has been multiplied by an overall amplitude factor. Besides this amplitude adjustment, the theory templates are in good agreement with mock results.

\begin{figure}
\centering
\includegraphics[width=\linewidth]{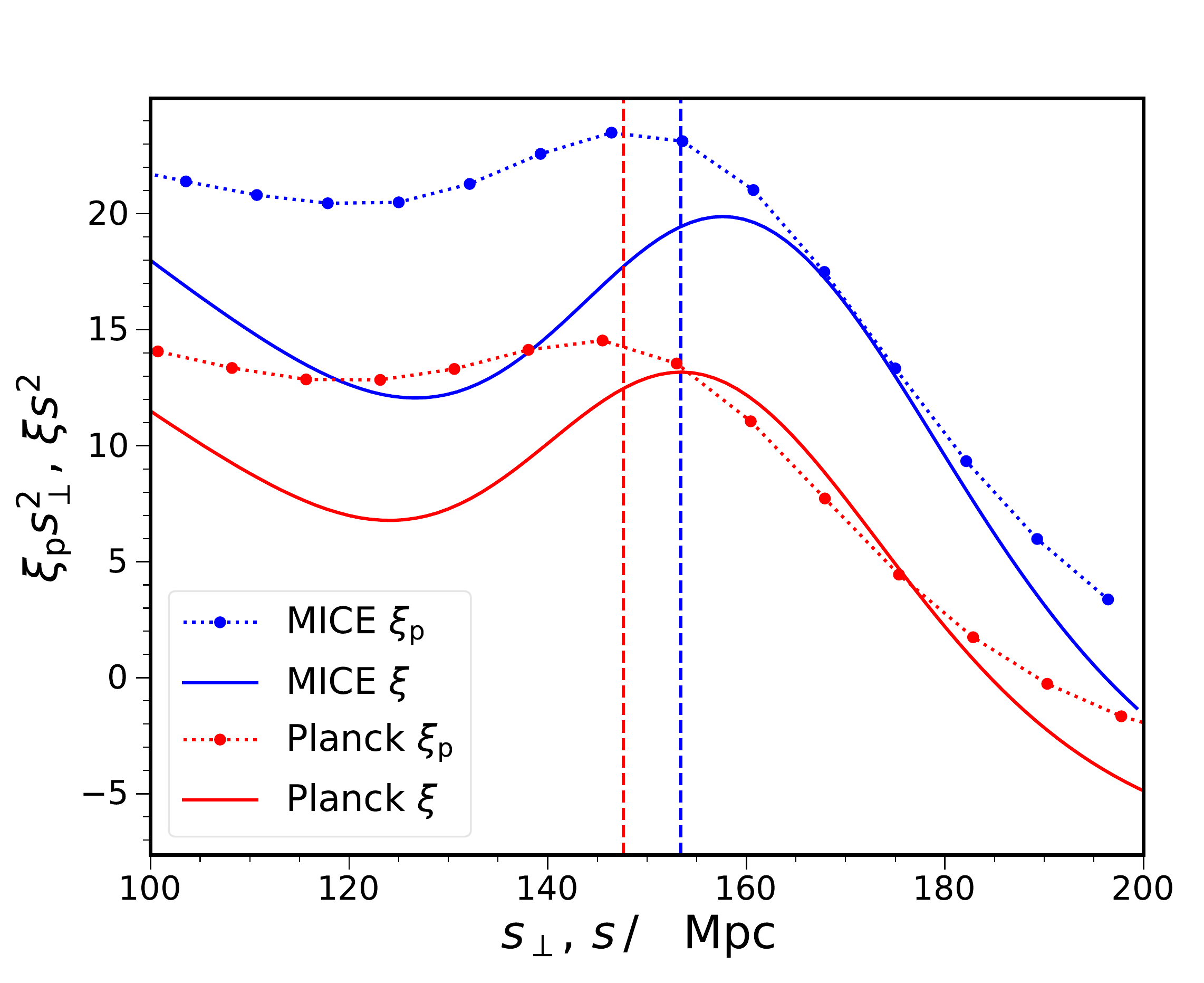}
\caption{ The photometric correlation function $\xi_{\rm p}$ versus $s_\perp $ (dotted line with filled circles) and the dark matter real space spectroscopic correlation $\xi$ against $s$ (solid) in MICE (blue) and Planck (red) cosmology, respectively. The sound horizons at the drag epoch in these cosmologies are also overplotted as dashed verticle lines (in the same color code). To enhance the BAO feature, they have been multiplied by the appropriate distance squared.   }
\label{fig:xi_spec_photo_BAO_cosmo}
\end{figure}

\change{ In the conventional angular tomographic analysis, the BAO features manifest in different angular scales, it is less direct to see the photo-$z$ effect on the BAO. To take advantage of the fact that the BAO information is condensed to a single bump in  $\xi_{\rm p} $, we plot in Fig.~\ref{fig:xi_spec_photo_BAO_cosmo} the photometric correlation $\xi_{\rm p} $ against $ s_\perp $ and the spectroscopic $ \xi $ against $s$. We have shown the results for both MICE and Planck cosmology.   The spectroscopic correlation $\xi $  is the dark matter real space correlation with isotropic BAO damping, obtained by an inverse Fourier transform of an analog of Eq.~\eqref{eq:Pk_anisotropic_damp} at  $z_{\rm eff} = 0.83  $. We also show the sound horizon  $r_{\mathrm{d}} $ in these two cases for comparison. Relative to the spectroscopic correlation, the BAO feature in $\xi_{\rm p} $ is less strong. Note that the BAO peak does not precisely correspond to $ r_{\rm d} $, and it is on different sides relative to the peak for the spectroscopic and the photometric cases.   A small bias could arise because of the subtle shape differences between the these two cosmologies. For example, the broadband terms introduced to take care of the shape difference, if their contribution is rapidly changing, it may slightly shift the BAO position. This is more likely to occur in the photometric case given the features are less sharp than in the spectroscopic one.  }

\begin{figure}
\centering
\includegraphics[width=\linewidth]{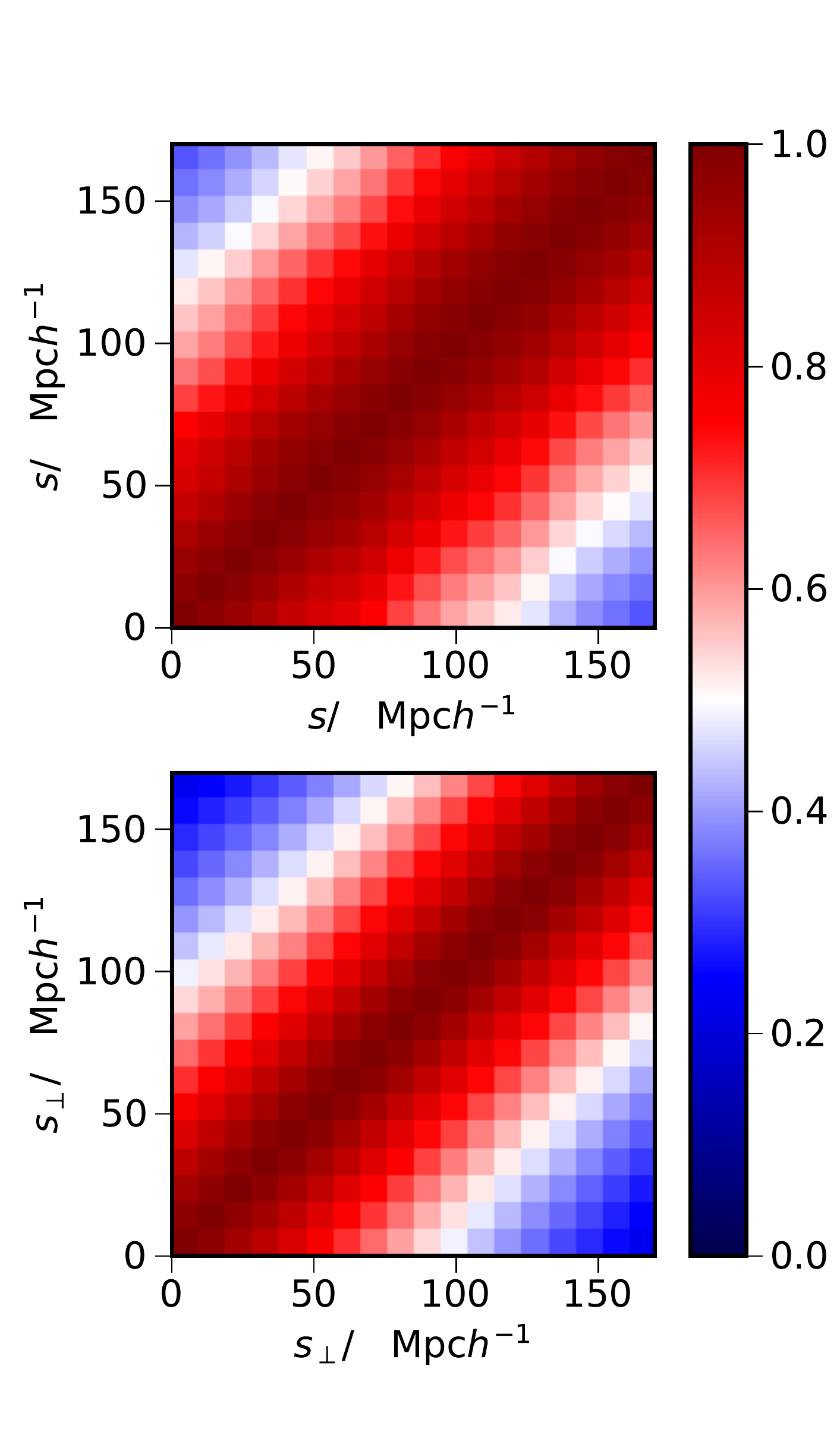}
\caption{ The correlation coefficient for $\xi_{\rm p}$. The upper panel shows the results for $s$ and the lower one for $s_\perp$. The results are for one $\mu $ bin in the range of [0, 0.8].   } 
\label{fig:corrcoef_Gauss_xi}
\end{figure}

We now turn to the covariance. In Fig.~\ref{fig:corrcoef_Gauss_xi}, we show  the correlation coefficient matrix for $\xi_{\rm p}$
\beq
\label{eq:correlation_coef}
\rho_{ij} \equiv \frac{ C_{ij} }{ \sqrt{ C_{ii} C_{jj} } }, 
\eeq
where $C_{ij} $ is an element of the covariance matrix.  Both the results for $s$ and $s_\perp$ are shown. In this plot we have used the Gaussian covariance. We have considered a single $ \mu$  bin in the range of [0,0.8].   These plots demonstrate that the covariance of $\xi_{\rm p} $ has strong off-diagonal elements, especially for  $s$. \change{  Real space results are known to be more correlated than the Fourier space ones. We argue that the situation for $\xi_{\rm p}$ is exacerbated by the photo-$z$ mixing. }

\begin{figure}
\centering
\includegraphics[width=\linewidth]{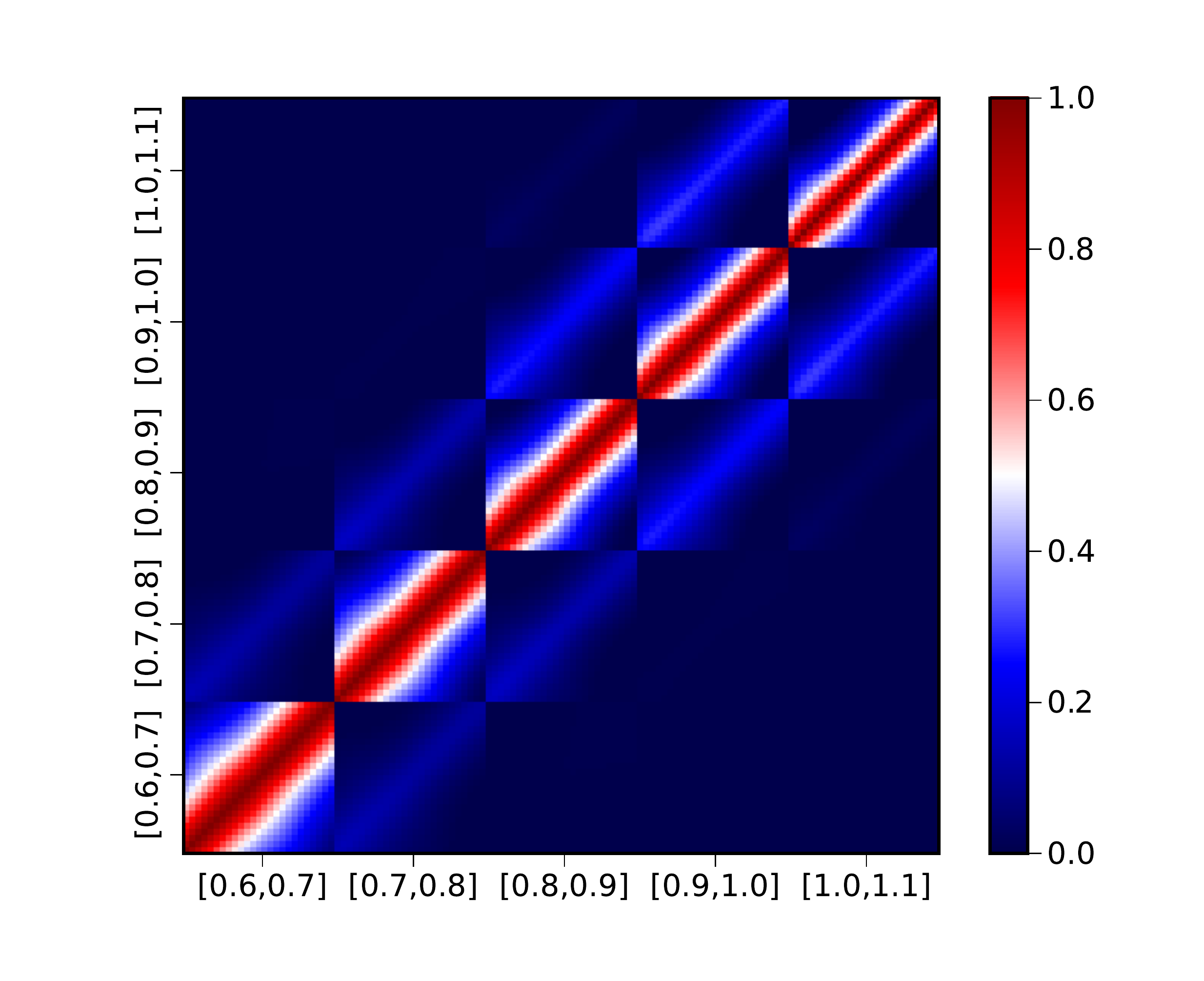}
\caption{ The correlation coefficient for $w$. The data in the redshift range [0.6,1,1] is split into five tomographic bins with bin size 0.1.  } 
\label{fig:corrcoef_Gauss_w}
\end{figure}

It is instructive to contrast the results with the Gaussian covariance for the angular correlation function, which is plotted in Fig.~\ref{fig:corrcoef_Gauss_w}. The setup of the calculations  are the same as those in the fiducial DES Y3 analyses.  There are five photo-$z$ bins, and in each bin, $\theta $ ranges from 0.5 to 5 degree with angular bin width of 0.2 degree. The photo-$z$ mixes modes in the radial direction.  Because the width of the redshift bin ($\Delta z = 0.1 $) is wide compared to the photo-$z$ uncertainty $\sigma $ [$\sim 0.03(1+z) $], the  photo-$z$ mixing does not affect the covariance in the same bin much.  The cross correlation between different redshift bins is mainly caused by photo-$z$ mixing, and the only significant correlation is between neighboring redshift bins.  Again because the redshift bin is wide compared to $\sigma $, the mixing between the neighboring bins is mild, and so even the neighboring bin covariance is weak.   On the other hand, as photo-$z$ mixing scale can be as high as $\sim 100 \MpcOh $, for $\xi_{\rm p} $ there can be substantial mixing between the underlying correlation from different scales. Thus the covariance for $ \xi_{\rm p} $ can be significantly higher than that for $w$.

As we mentioned, construction of the ICE-COLA mock catalogs involves replicating the simulation box to form a full-sky light-cone mock. The repeated structures cause the covariance to be anomalously higher than the correct one, and this prevents us from using the mock covariance as the benchmark to test our Gaussian theory covariance.   See \citet{COLAmocks_Ismael} for more discussions. Thus in DES Y3, the analytic Gaussian covariance is the default choice. Although we cannot directly test the $\xi_{\rm p}$ covariance against mock results, the code used to compute the Gaussian covariance for $w$, which is the backbone of the $\xi_{\rm p} $ results, has been tested in \citet{Chan:2018gtc,Avila:2017nyy,y3-baomain}.  Moreover, we have verified that the amplitude ratio difference between the Gaussian theory covariance and the mock covariance are at a similar level for $w$ and $\xi_{\rm p}$, albeit quantitative comparison is not possible.





\subsection{BAO fit} 
\label{sec:BAO_fit}

In this subsection, we apply our theory template and covariance to extract the BAO scale from the mock catalogs.  With the variable $s_\perp$, the BAO scale is almost equal to the true one, we adopt $\xi_{\rm p}$  as a function of $s_\perp $ in the BAO fit.  To measure the BAO scale, as in standard BAO analyses [e.g.~\citet{Xu_etal2012,Anderson_etal2014,Ross:2017emc}], we apply the following general model
\beq
\label{eq:full_nuisance_model}
M(s_\perp) = B \, T( \alpha s_\perp )  + \sum_i \frac{ A_i }{ s_{\perp}^i },     
\eeq
where $T$ denotes the theory template and $B$ gives the freedom to adjust the amplitude. The parameter $\alpha$ in the argument of $T$ is the key parameter we are after. It allows us to shift the BAO scale computed in the fiducial cosmology to match that in the data.    The polynomial in $1 / s_\perp $ is designed to absorb residual broadband effects due to imperfectness in modeling, changes in correlation function shape as a result of the differences between the fiducial cosmology and the actual one, and residual broadband systematic effects.

The default configuration for the BAO fit is as follows: the broadband terms consisting of $\sum_i A_i / s_\perp^i$ with $ i=0$, 1, and 2; the fit range  $[40,140] \MpcOh$; the bin width $\Delta s = 3 \MpcOh $, and the Gaussian theory covariance.  We use 384 mocks for the BAO fit test.   Different configurations for BAO fit are checked.

We look for the best fit $\alpha $ by the maximum likelihood estimator following the procedures in \citet{Chan:2018gtc}. We first analytically fit the linear parameter $A_i$, and then minimize the resultant $\chi^2$ w.r.t. $B$ subject to the constraint $B>0$. At the end of the operations, we are left with the residual $\chi^2$ as a function of $\alpha$. The best fit $\alpha$ corresponds to the point where the residual $\chi^2 $ attains its minimum and the symmetric 1-$\sigma $ error bar is derived from the $\Delta \chi^2 =1$  criterion. Besides we also cross check our results with the Markov Chain Monte Carlo (MCMC) method, in which all the parameters are fitted simultaneously.

\subsubsection{ Highly correlated data fit } 
\label{sec:highly_correlated_fit}
The photo-$z$ mixing causes $\xi_{\rm p} $  to be strongly correlated, as evident in Fig.~\ref{fig:corrcoef_Gauss_xi}.  Here we first describe some fitting problems caused by the high correlation, namely the apparently bad fit results and the alarmingly large $\chi^2 $ per degree of freedom ($\chi^2 /{\rm dof} $). The former issue is caused by the largest eigenvalues of the covariance and the latter is due to the small ones. We discuss a few approaches to modify the covariance, and find that the first issue can be resolved by suppressing the largest eigenvalues. For the second issue, while it is not practicle to solve it by adjusting the eigenvalues, we demonstrate that the BAO feature is well-fit by our model and the problem is not directly related to the BAO.

\begin{figure}
\centering
\includegraphics[width=\linewidth]{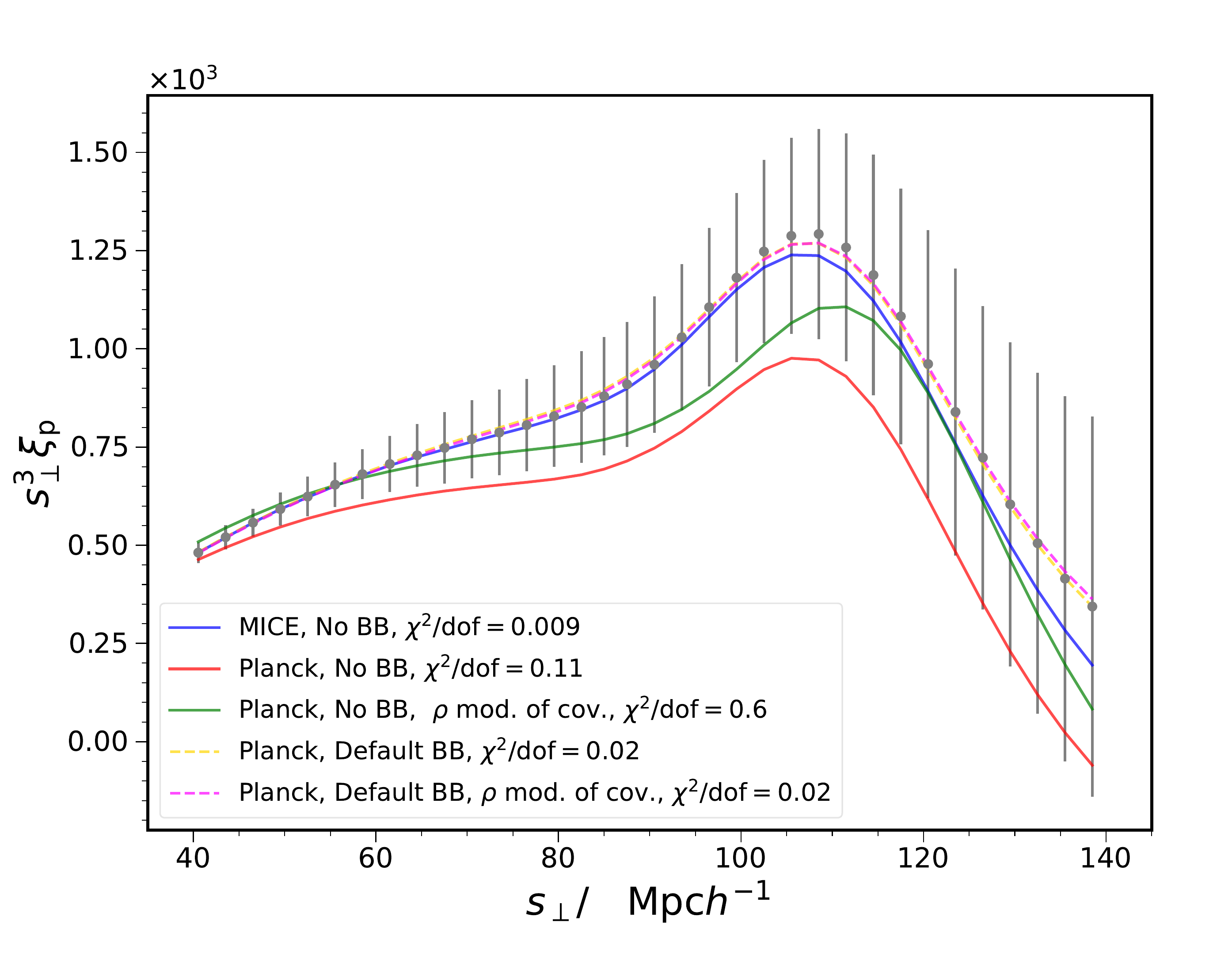}
\caption{ The BAO fit results to the mean of the mock measurements (grey data points with error bars). The fit obtained with the MICE template without broadband terms (solid, red) is compared with the Planck template without broadband terms result (solid, green). The latter case is below all the data points.  With the inclusion of the broadband terms, the Planck template (dashed, yellow) also offers good fit.  The results for the modified covariance are also shown [No broadband terms (solid, green) and with broadband terms (dashed, violet)].  }
\label{fig:peculiar_meanfit}
\end{figure}

To begin with, we present some peculiar fit results.  One of the odd examples appears in the BAO fit to the mean of the mocks, which are shown in Fig.~\ref{fig:peculiar_meanfit}.  When the MICE cosmology template is used, the resultant fit is very good and it is not surprising that the $\chi^2 /{\rm dof} $  is very small (0.009).  However, when the template computed in Planck cosmology  \citep{Planck2020} is used, we find that the resultant $\chi^2  /{\rm dof} $  is still small (0.11), but the fit is manifestly poor by eye because the best fit model is below {\it all} the data points and most of the data points are more than  1-$\sigma $ away from the best fit.

Although this issue only appears when the model Eq.~\eqref{eq:full_nuisance_model} does not include the broadband terms at all and disappears otherwise, it does indicate that something abnormal happens to the fit \footnote{The issue is not caused by the fitting method.  Usage of the MCMC method also gives the same results. }.    In the other extreme, when we fit to the individual mocks, we find that the fit results are often fine by eye as the best fit model fall within all the 1-$\sigma$ errors, but  $\chi^2 /{\rm dof} $ is alarmingly large, some of them reach as high as 2.  Note that the second issue we describe here appears for generic fit configurations, in particular for both Gaussian and mock covariance fit.  We point out that the second problem is already implicit in \citet{Abbott:2017wcz} and \citet{Sridhar_etal2020}, as their  $ \xi_{\rm p} $  fit results look decent by eye but appreciably large  $\chi^2 /{\rm dof} $ were reported.  We attribute these problems to the high correlation of the data induced by the photo-$z$ mixing.

\begin{figure}
\centering
\includegraphics[width=\linewidth]{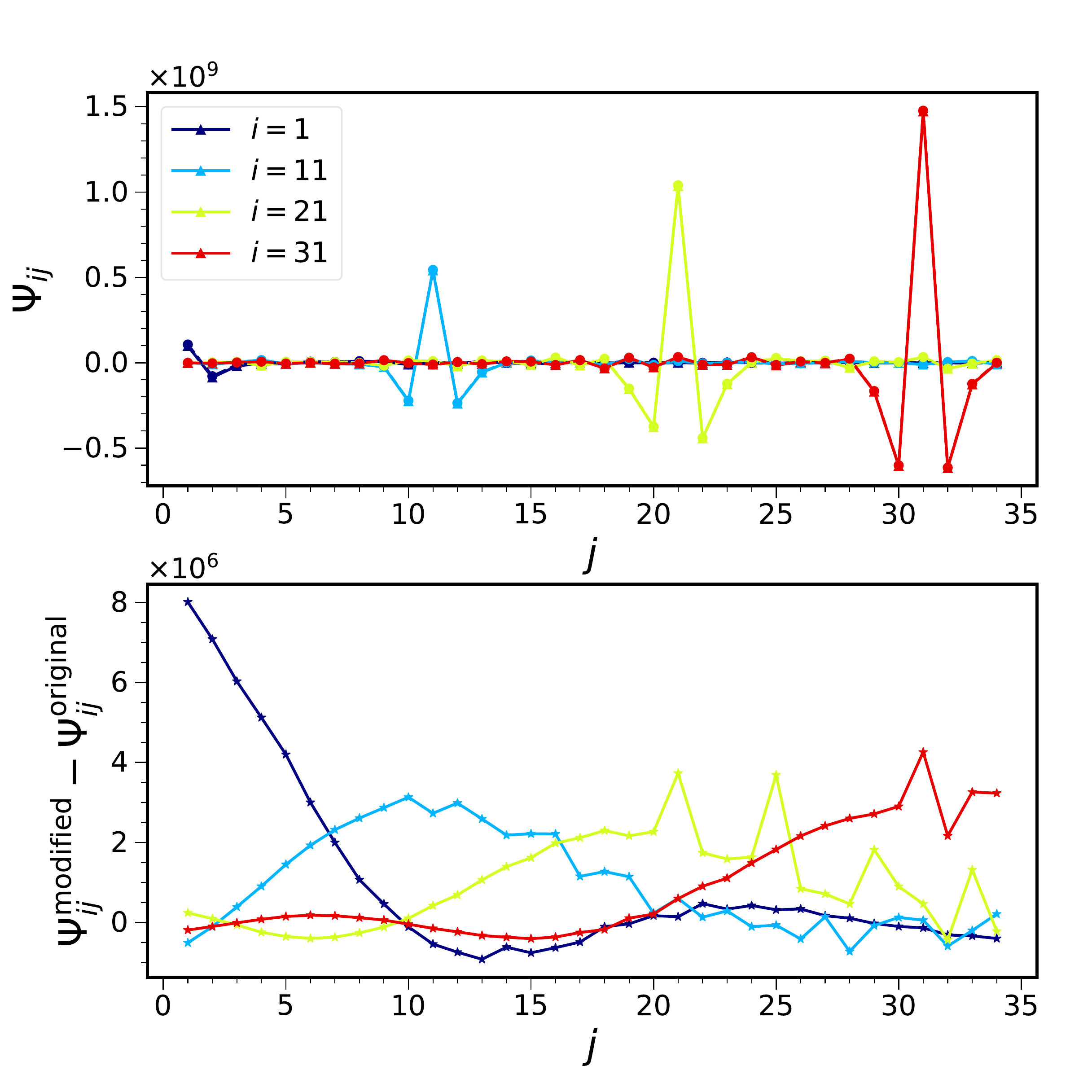}
\caption{  Upper panel: The elements of the precision matrix $\Psi_{ij} $ plotted against $j$ for fixed $i$. The elements for the original $\Psi $ (triangles) and the one obtained by modifying the covariance matrix (circles, see the text for details) are compared.  Lower panel: The differences between the modified  $\Psi_{ij} $ and the original one are displayed. The effect of the modified covariance is mainly to shift $\Psi $ to the positive side.  As the large eigenvalue modes correpsond to eigenvectors of long ``wavelength'', it also weakly affects the zones far from the diagonal.    }
\label{fig:Psi_ij_elements}
\end{figure}

Although the covariance of $\xi_{\rm p}$ has strong off-diagonal elements, its inverse, the precision matrix $\Psi$ could be approximated by a band matrix with a few non-vanishing elements near the diagonal. The upper panel of Fig.~\ref{fig:Psi_ij_elements} shows the elements of $\Psi$, and it is approximately a pentadiagonal band matrix, i.e.~a band matrix has non-zero elements only for the main diagonal, and the first two upper and two lower diagonals.  We can use this to intuitively understand why the best fit may not be driven to the data points as close as possible.  By assuming $\Psi $ to be pentadiagonal, the $\chi^2 $ can be written as
\begin{align}
  \chi^2  = \sum_{j=1}^N   &\big(  \Delta_{j-2} \Psi_{j-2,j} +  \Delta_{j-1} \Psi_{j-1,j} +  \Delta_{j} \Psi_{j,j} \nn \\
         & + \Delta_{j+1} \Psi_{j+1, j} +  \Delta_{j+2} \Psi_{j+2, j} \big) \Delta_j ,  
\end{align}  
where $ \Delta $ denotes the difference between the model and the data. This expression can accomodate the results near the ``boundaries'' as well if   $\Psi_{i,j}\equiv 0 $ whenever $i, \, j <1$ or $>N$. Now suppose $ \Delta_j>0 $, because  $\Psi_{j,j}>0$, to minimize $\chi^2$, $\Delta_j $ is attracted to zero. For diagonal covariance, this is the sole contribution. However, because  $\Psi_{j-2}$, $\Psi_{j-1}$, $\Psi_{j+1} $, and $\Psi_{j+2} $ are negative,  it is favorable for $ \Delta_{j-2}$,  $\Delta_{j-1}$, $ \Delta_{j+1}$, and  $\Delta_{j+2}$ to be repelled to as negative as possible. Of course these expressions are coupled, more careful analysis is required,  but it is likely that the competition between the diagonal terms and the off-diagonal ones lead to a best fit that does not precisely fall on the data.

\begin{figure}
\centering
\includegraphics[width=\linewidth]{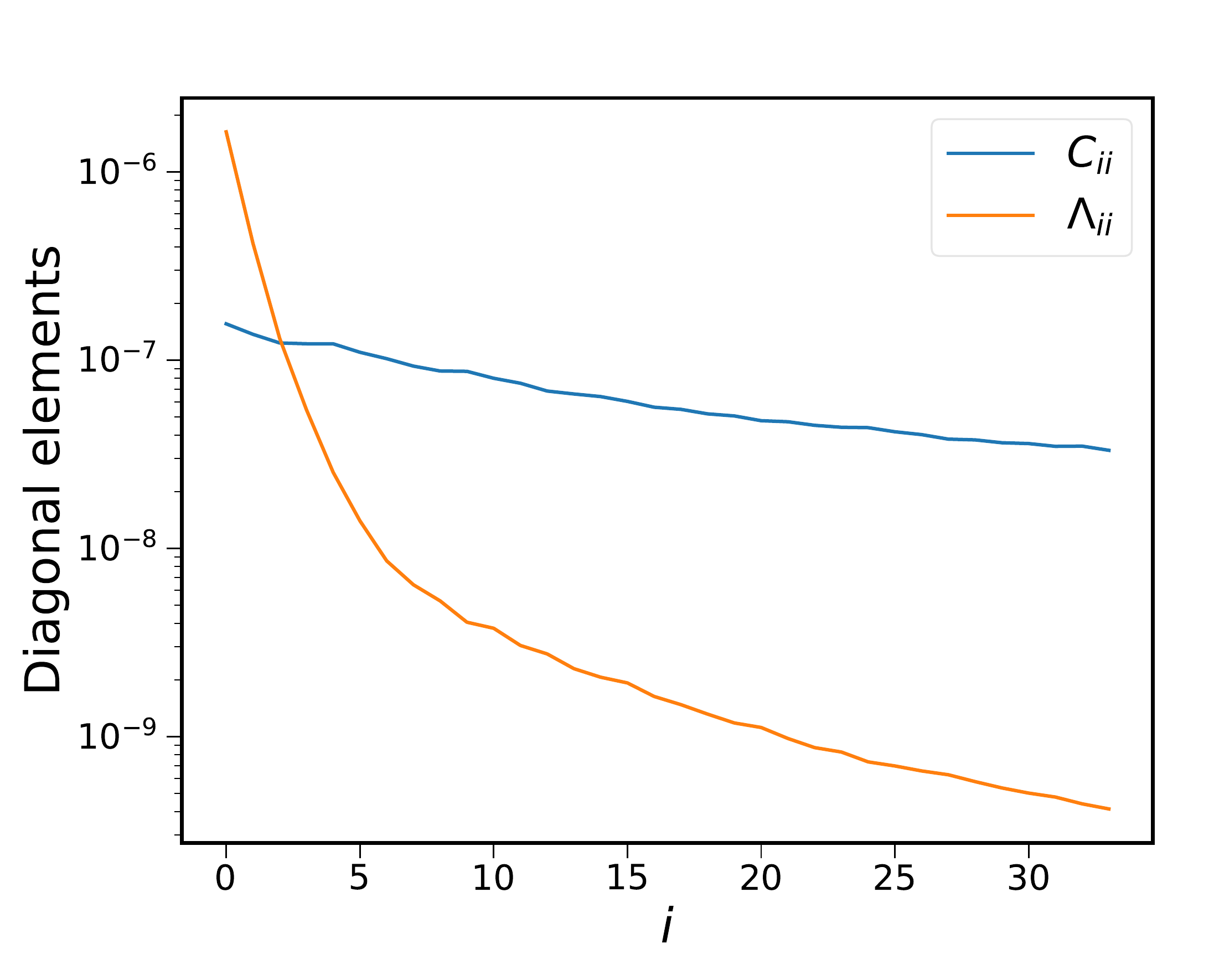}
\caption{ The eigenvalue spectrum of the covariance matrix (orange) and the diagonal elements of covariance matrix (blue). While $C_{ii}$ is in its natural order, the eigenvalues have been aranged in descending order.    }
\label{fig:Diagonal_C_Lambda}
\end{figure}

To look into the covariance, we plot its eigenvalues in Fig.~\ref{fig:Diagonal_C_Lambda}.  We find that the spectrum spans three orders of magnitude, and most of them are of relatively small values.  As a comparison we also show the diagonal elements of $C$, its range is much more limited.  For the fit to the mean of mocks using the Planck template, we find that the fit is dominated by the largest eigenvalue mode. Because of its (excessively) large eigenvalue, its contribution to the $\chi^2$ is very small, so that even the fit is visually bad, $\chi^2 /{\rm dof} $ is still small. On the other hand, for the fit to individual mocks, the large $\chi^2 /{\rm dof}$ values are caused by significant contributions from modes with small eigenvalues.

The problem of fitting highly correlated data is also encountered in the lattice field theory because lattice simulation data are highly correlated. \citet{Jang:2011fp} surveyed a few methods proposed by the lattice field theory community to tackle this problem. \cite{Michael:1993yj} advocated ignoring the off-diagonal elements in the covariance, but this is clearly a too violent modification.  A more mild method is to remove the eigenmodes deemed problematic. Both removing the largest eigenvalue modes \citep{Thacker:1990bm} and the smallest ones \citep{Drummond:1992pg,Kilcup:1993ur} have been proposed. For the largest eigenvalue modes removal, it was argued that these modes have small contribution to $\chi^2 $. The smallest eigenvalue modes removal is often favored in the spirit of the singular value decomposition \citep{Press:2007:NRE:1403886}. A physical argument in support of it is that  those modes are generally highly oscillatory and so they are not conducive to getting a smooth fit.   In a similar vein, \citet{Bernard:2002pc} proposed to modify instead the correlation coefficient Eq.~\eqref{eq:correlation_coef}. They replace the eigenvalues of $ \rho $ smaller than  certain threshold $\lambda_{\rm th} $  by   $\lambda_{\rm th} $. A new covariance matrix can be constructed  using the original eigenvectors and the modified eigenvalues.  Modifying the eigenmodes is more attractive as it keeps all the directions in the covariance space constrained.

\begin{figure}
\centering
\includegraphics[width=\linewidth]{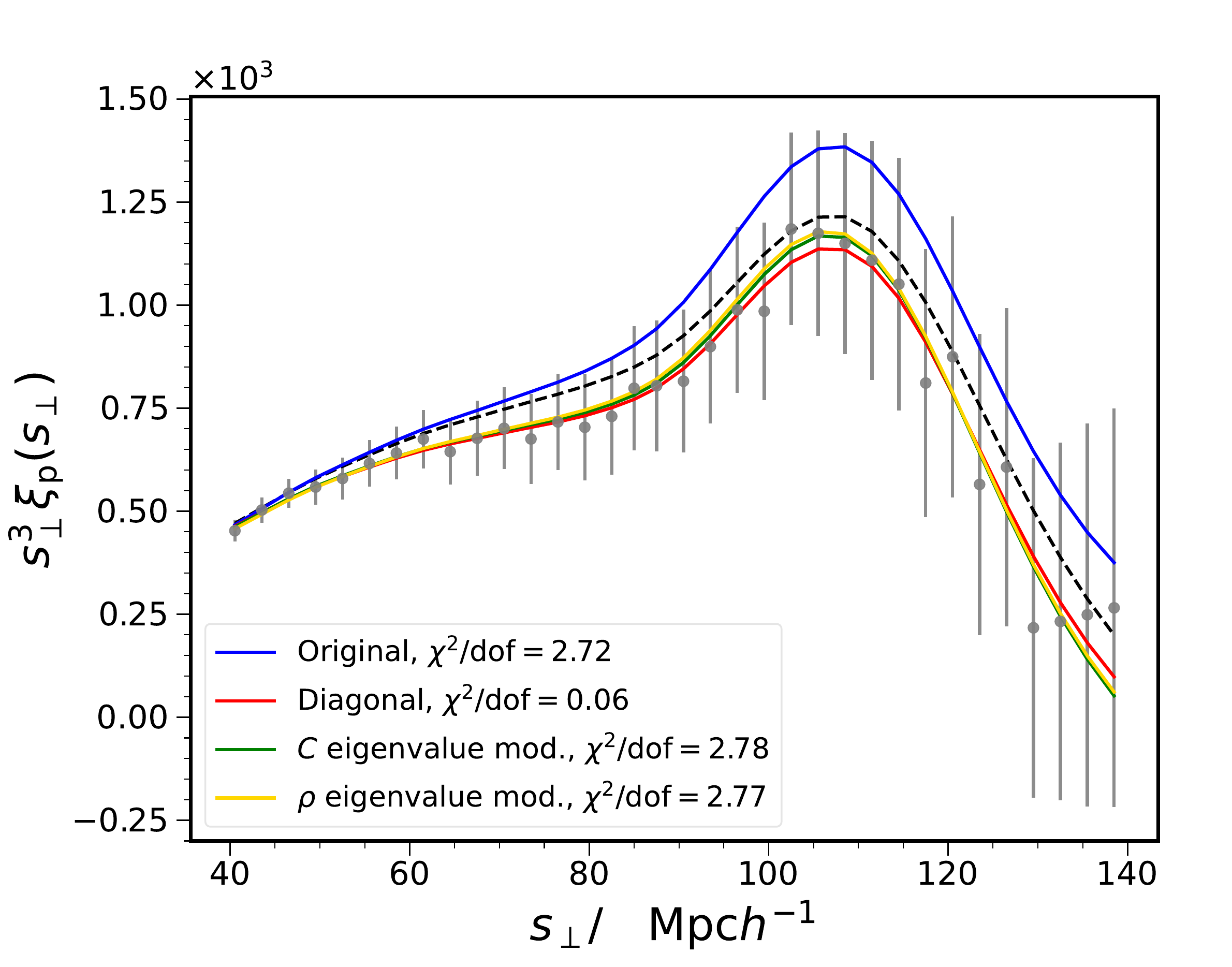}
\caption{ The best fit to the $\xi_{\rm p} $ measurement from one of the mocks (circles, the error bars are 1-$\sigma$ error from the diagonal elements of the covariance). This mock is chosen because the differences among different covariance recipes are pronounced. The results obtained with the original covariance matrix (blue), diagonal approximation (red), suppressing the largest eigenvalues in the covariance (green), and suppressing the largest eigenvalues in the correlation coefficient (yellow) are compared. The raw template (black, dashed) is also plotted.   }
\label{fig:xip_fit_Cmod_compare}
\end{figure}

We test the results obtained with different covariance recipes.   While for most of the fit, different approaches give similar results, in a small fraction (order of 10\%) the differences are quite pronounced. In Fig.~\ref{fig:xip_fit_Cmod_compare}, we show the fit results from one of the mocks, in which the differences are clearly visible. The one obtained with the original covariance (blue) is manifestly not optimal. The best fit is  $ 1.006 \pm  0.017 $ in this case.  In the other extreme, when the diagonal approximation (red) is used, it is not surprising that the best fit is driven close to data points, but the best fit we get is  $1.002 \pm 0.032 $, with the error bar substantially larger than the original one.  By suppressing the largest eigenvalues of the covariance (green) or $\rho$ (yellow), we can also get an apparently better fit.  We replace the eigenvalues larger than certain thereshold $ \lambda_{\rm th }  $  by $ \lambda_{\rm th }  $.  We take $ \lambda_{\rm th}$ to be  $ 3 \times 10^{-8} $ and 1 for the covariance and correlation coefficient modification respectively. The thresholds are chosen so that as few eigenvalues are affected as possible and the affected ones account for about 10\% of them.   The best fits we get are $1.003 \pm 0.016  $ and $1.005 \pm 0.017   $ respectively.

In Fig.~\ref{fig:Psi_ij_elements} we also show the elements of $\Psi$ obtained by suppressing the largest eigenvalues of $\rho$. Overall, the size of the modification is small, of the order $10^{-3} $.  The main effect of the covariance modification is to shift $\Psi $ to the positive side  so that  the attraction to zero is enhanced and the repulsion to negative infinity is weakened. Note that the large eigenvalue modes correspond to eigenvectors that are oscillatory in nature with long ``wavelength'', while the small eigenvalue ones tend to be well localized.  Thus the modification also weakly affects the the zone far from the diagonal.     Furthermore, we find that the $\rho $ modification recipe often results in better agreement with the original one compared to the covariance modification, hence we will adopt $\rho $ modification in the following analysis.  In Fig.~\ref{fig:peculiar_meanfit}, we also show the results obtained with  $\rho $ modification. Without the broadband terms, the best fit now passes through the data points, although the BAO position appears to be biased. With the default braodband terms ($A_i$ with $i=0$, 1, 2, and 3), the bias is removed.

On the other hand, the large $\chi^2 / {\rm dof}  $ problem is caused by the small eigenvalues. The nominal $\chi^2/{\rm dof} $ for the example shown in Fig.~\ref{fig:xip_fit_Cmod_compare} are similar for different recipes, around $~2.7$, except for the diagonal approximation, which gives  $\chi^2/{\rm dof} \sim  0.06 $. Since only the largest eigenvalues are modified, the $\chi^2 $ is only increased slightly compared to the original one.  As it is clear from the spectrum in Fig.~\ref{fig:Diagonal_C_Lambda}, most of the eigenvalues are small and of similar values, we need to remove or modify a large fraction of eigenmodes before the effect becomes noticeable.  Removing 40\% of the eigenmodes with the smallest eigenvalues, the naive $\chi^2 / {\rm dof} $ can be reduced to roughly 1. This represents a very substantial modification of the covariance matrix. Moreover, after eigenmodes removal or modification, there is no proper theoretical basis for $\chi^2 / {\rm dof} $ anymore \citep{Bernard:2002pc}. Thus we shall not pursue to reduce  $\chi^2 / {\rm dof} $ by adjusting the covariance matrix.

\begin{figure}
\centering
\includegraphics[width=\linewidth]{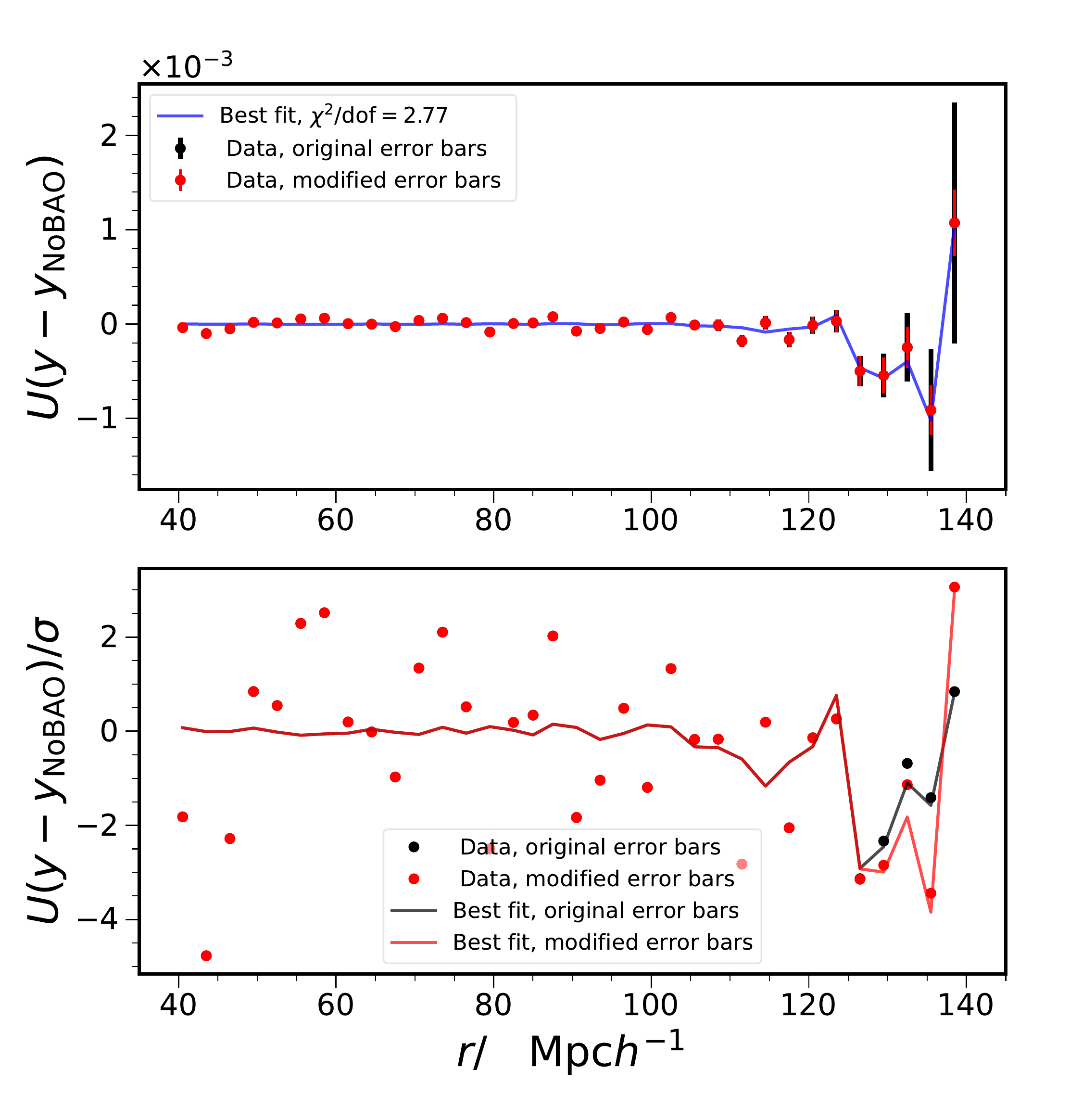}
\caption{ Upper panel: The data (filled circles) and the best-fit model (solid blue) shown in the rotated basis. The orthogonal matrix $U$ decorrelates the correlation in the error.   To highlight the BAO feature, the no-BAO template has been subtracted. In this basis, the BAO feature manifests as a series of wiggles for $r \gtrsim 110 \MpcOh $.  The uncorrelated error bars from the original covariance (black) and the modified covariance obtained from the $\rho$ modification scheme (red) are displayed.  The modified error bars are reduced compared to original ones only in the large $r$ range.  The data used in this plot is based on the same mock data used in Fig.~\ref{fig:xip_fit_Cmod_compare}.  Lower panel: The results are further divided by the error bars to show the significance of the deviations between the model and the data points. The results obtained with the original error bars (black) and the modified error bars (red) overlap with each other and deviations show up only for  $ r>130 \MpcOh $. In the BAO zone, the model and the data points agree with each other well. The large $ \chi^2 / {\rm dof} $ is driven by the data points lying outside the BAO feature.  }
\label{fig:Uy_NoBAO}
\end{figure}

Using the example in Fig.~\ref{fig:xip_fit_Cmod_compare} again, we check directly if the BAO scale is well fitted by the model in Fig.~\ref{fig:Uy_NoBAO}. To decorrelate the data, we rotate the data vector by the matrix $ U$, which is the orthogonal matrix that diagonalizes the covariance, i.e.~$C=U^{T} \Lambda U$.  To highlight the BAO feature in the rotated coordinates, the no-BAO template has been subtracted from both the data vector and the best fit model. Note that as the data vector is mixed by $U$, they on longer strictly correspond to the original distance $r$.  Here we only use it as a convenient label for the $x$-axis.  Nonetheless, from the difference with the no-BAO template, the location of the BAO feature can be inferred.   In this basis, the BAO feature shows up as a series of wiggles in $ r \gtrsim 110 \MpcOh $ instead of a single peak in the usual configuration representation. The error bars on the data are given by the eigenvalues in $\Lambda$ and they are not correlated in this basis.  We have also shown error bars from the modified covariance according to the $\rho$ modification recipe. We see that the modification reduces the error bars around the BAO scale. 
To highlight the source of large $\chi^2 /{\rm dof} $, we show the ratio  with respect to the error bar in the lower panel of  Fig.~\ref{fig:Uy_NoBAO}.  Both the results obtained with the original error bars and the modified ones are shown, which coincide with each other except for $ r > 130 \MpcOh $.   Significant deviations of the best fit model from the data points occur for $r\lesssim 100 \MpcOh$,  while in the BAO zone, the model and the data agree with each other well for both types of error bars.  Thus we conclude that the large  $\chi^2 /{\rm dof} $  is caused by the part of the fit that is not directly related to the BAO.

\begin{table*}
  \caption{  BAO fit results on the COLA mocks with different configurations in the analysis. The default setup is as follows: the broadband terms consisting of $\sum_i A_i / s_\perp^i$ with $ i=0$, 1, and 2, the fit range  $[40,140] \MpcOh$, the bin width $\Delta s = 3 \MpcOh $, and  Gaussian theory covariance with eigenvalues of  $\rho$  suppressed.  See the texts for details.  }     
  \label{tab:BAO_mocktest}
  \begin{tabular}{lccccccccc}
  \hline
  \hline
  case &  $\langle \alpha \rangle$     &   $\sigma_{\rm std}$   &   $\sigma_{68}$      & $\langle \sigma_{\alpha} \rangle$    &  ${\rm frac.\,encl.}   \langle \alpha \rangle$   &  $\langle d_{\rm norm} \rangle$   & $\sigma_{d_{\rm norm}}$  &   ${\rm mean\,of\,mocks}$     &  $\langle \chi^2 \rangle / {\rm dof}$       \T\B  \\  \hline 
  Default                        &  1.001   &  0.023  & 0.022  & 0.020   &  61\%    & -0.005      &  1.18    &  1.002 $\pm$ 0.019     &  58.7/29 (2.03)    \\ \hline  
   $w$ fit           &  1.004   &  0.024  & 0.023  & 0.021   &  62\%    & -0.024      &  1.11    &  1.004 $\pm$ 0.021     &  93.1/89 (1.05)    \\ \hline  
$A_i, \, i=0$                  &  1.001   &  0.023  & 0.022  & 0.019   &  59\%    &  0.032      &  1.19    &  1.002 $\pm$ 0.019     &  62.9/31 (2.03)    \\ 
$A_i, \, i=0,1$                &  1.001   &  0.023  & 0.021  & 0.021   &  64\%    &  0.003     &  1.12     &  1.003 $\pm$ 0.020     &  58.6/30 (1.95)    \\ 
$A_i, \, i=0,1,2,-1$           &  1.002   &  0.023  & 0.021  & 0.020   &  60\%    & -0.005     &  1.17     &  1.002 $\pm$ 0.019     &  56.8/28 (2.03)    \\  
\hline
Fit range $[70,130] \MpcOh$    & 1.000    &  0.027  & 0.024  & 0.022   &  59\%    & 0.084      &  1.20     &  1.003 $\pm $ 0.021   &  36.4/16 (2.28)     \\   
\hline
$\Delta s = 10 \MpcOh$         & 1.001   &  0.025   & 0.024  & 0.020    & 57\%     & -0.007    &  1.25     &  1.001 $\pm$ 0.020    &  10.2/5 (2.05)      \\  
$\Delta s = 5 \MpcOh $         & 1.003   &  0.025   & 0.023  & 0.020   & 59\%      & -0.022     &  1.17    &  1.002 $\pm$ 0.020    &   32.1/15 (2.14)     \\  
$\Delta s = 2 \MpcOh $         & 1.002   &  0.024  & 0.022  & 0.020   & 60\%      & 0.008      &  1.12     &  1.002 $\pm$ 0.019    &  81.0/45 (1.80)      \\   
$\Delta s = 1 \MpcOh $         & 1.001   &  0.025  & 0.023  & 0.019   & 57\%      & 0.019     &  1.37      &  1.002 $\pm$ 0.018    &  143.1/95 (1.51)      \\  
\hline
Planck template                &  0.958 &  0.022  & 0.020  & 0.019   & 65\%      & -0.003     & 1.14       &  0.957 $\pm$ 0.018    &  53.7/29 (1.85)      \\  
\hline    
Original Gaussian cov.         &  1.002   &  0.023  & 0.022  & 0.021   &  64\%    & -0.005     &  1.13    &  1.003 $\pm$ 0.020     &  57.7/29 (2.00)     \\ 
COLA cov                       & 1.001   &  0.023  &  0.022 &  0.021  & 65\%      & -0.009     & 1.16      & 1.002$\pm$ 0.021      &  47.9/29 (1.65)      \\  
\hline
MCMC fit                       & 1.001   & 0.024   & 0.023  & 0.020   & 59\%      & -0.004     & 1.18      & 1.003$ \pm$ 0.020     &        \\  
\hline
$ N_{\mu} = 3 $                 & 0.999   & 0.028   & 0.025  & 0.017   & 47\%      & -0.024     & 1.60       & 0.999$\pm$ 0.017     &  139.0/89 (1.56)   \\  
$ N_z  = 2  $                  & 1.000   & 0.023   & 0.022  & 0.018   & 59\%      & -0.014     & 1.24      & 1.001$ \pm$ 0.018     &  98.1 /59 (1.66)    \\   
\hline \hline
  \end{tabular}
\end{table*}

\subsubsection{ Fit results } 
\label{sec:BAOfit_results}

In this section, we present the BAO fit results in details. We will consider various metrics to quantify the accuracy of the fit and different variation of the fit configurations. Although these are important, they can be tedious to read. The key default result is given in the first line in Table~\ref{tab:BAO_mocktest}. In particular, the $\xi_{\rm p } $ constraint on $\alpha$ is  $\langle \alpha  \rangle \pm  \sigma_{\rm std} =  1.001 \pm 0.023 $.  We have checked that the results are robust against variation of the fit configurations.  The default setup uses the Gaussian covariance modified with the $\rho$ modification prescription.  We shall also contrast them with the corresponding results for the angular correlation.   They are also consistent with the $w$ results even though  $\xi_{\rm p } $ statistic has considerably larger  $ \langle \chi^2 \rangle / {\rm dof} $.


The default case results for $\xip$ fit are shown in the first row in Table~\ref{tab:BAO_mocktest}. As a reference, we also show the corresponding results for the angular correlation function, which are obtained from a joint five-tomographic fit. See Table II in \citet{y3-baomain} for more details on the $w$ fit.   The mean of the best fit $\langle \alpha \rangle$ is $1.001$, with small bias which can be attributed to nonlinear evolution \citep{CrocceScoccimarro_2008,PadmanabhanWhite_2009}.  Two measures of the spread of the distribution of the best fit $\alpha$ are shown: the standard deviation $\sigma_{\rm std} $ and the half of the width between 16 and 84 percentile of the distribution,  $\sigma_{68} $, which is less sensitive to the tails of the distribution.  We find that $\sigma_{\rm std}$  is larger than  $\sigma_{68 } $ by a small amount (0.001), suggesting the influence by the tails.   As mentioned, the  1-$\sigma$ error bar is derived from the $\Delta \chi^{2}=1$ condition, and for the error bars to be meaningful, $\langle\sigma_\alpha \rangle$ should agree with the measures of the spread of the distribution. In the case of Gaussian likelihood, they are equal to each other. We find that the mean of the error bar is slightly smaller than the spread of the distribution. These numbers are marginally better than those for $w$. The bias in the best fit  $\langle \alpha \rangle $ is smaller (1.004 for $w$) and the measures of spread and error are smaller by $\sim 0.001 $.  These (mild) improvements could arise from the fact that  the $\xi_{ \mathrm{p} }$ statistic makes use of the cross correlation in the data as well, not only the tomographic bin auto-correlation for $w$ as in the study of DES Y3.

Another indicator for the accuracy of 1-$\sigma$ error bar is the fraction of times that it encloses $\langle\alpha\rangle$, which is $68 \%$ in the case of Gaussian likelihood.  In line with the error bar being slightly smaller than the spread, the fraction enclosing $\langle\alpha\rangle$ is slightly lower than the Gaussian expectation.   The normalized deviation, $d_{\text {norm }} \equiv ( \alpha - \langle \alpha \rangle ) / \sigma_{\alpha}$ is designed to study the correlation between the deviation of the best fit $\alpha$ from the ensemble mean and the error bar derived.   
The mean $\left\langle d_{\text {norm }} \right\rangle$ is close to zero, but the standard deviation  $\sigma_{d_{\text {norm }}}$ exceeds unity by $18 \%$. This trend is consistent with the previous observation that the spread of the distribution is slightly larger than the error estimate. In contrast, for $w$, the fraction of time enclosing $\langle \alpha \rangle$ is  62\%, and  $\left\langle d_{\text {norm }} \right\rangle = 0.02$ and $\sigma_{d_{\rm norm} } = 1.12 $. Both suggest that $\xi_{\rm p} $ results are less Gaussian than the $w$ ones.

As a reference, we also show the fit to the mean measurement of the mocks. To show the results for a ``typical'' fit, we have used the covariance corresponding to a single survey volume. The results are similar to $\langle\alpha\rangle \pm \left\langle\sigma_{\alpha}\right\rangle$. However, we note that for $w$ these different measurement are almost always equal to each other, suggesting that the averaging and fitting operations are more commutative for $w$.

Finally, the $ \langle \chi^2 \rangle / {\rm dof} \, (\sim 2)$ is also shown for reference,  but we should bear in mind the complications mentioned in the previous subsection. On the other hand, for $w$,  $ \langle \chi^2 \rangle / {\rm dof} $ is 1.05. We attribute this to the fact that the covariance for  $w$ is much less correlated.

A number of tests on the variation of the fitting configurations have been performed.  We have tested the number of the broadband terms $\sum_i A_i / s_\perp^i $ with $i=0$, $i=0,1$, and $i=0$, 1, 2, -1.  Besides minor fluctuations, the fit results are not sensitive to the number of broadband terms used. This is because the broadband terms are not degenerate with $\alpha$, but they are degenerate among themselves.  When a narrower fitting range $[70,130] \, \MpcOh$ is used, the constraining power is reduced and so the spread of the distribution is enlarged, especially for $\sigma_{\text {std }}$.  As the error bar $\sigma_{\alpha}$ increases only mildly, the difference between the error bar and the spread of the distribution increases.

The bin width  size $\Delta s = 10$, 5, 2, and 1 $\MpcOh$ are compared with the  fiducial setting $\Delta s = 3 \,  \MpcOh $. Among the cases shown, $\sigma_{\text {std }}$ for $\Delta s = 3 \,  \MpcOh $ is the lowest and it is also more consistent with $\sigma_{68} $ and  $\langle \sigma_\alpha \rangle $. Note that $\chi^2 /{\rm dof}$ are smaller for  $\Delta s = 2$ and 1 $\MpcOh$, but this is not vital here as we already mentioned that the BAO region is well-fit by our model, the reduction in $\chi^2 /{\rm dof}$ is primarily due to the region lying outside the BAO.

To test if the expected shift in $\alpha$ can be recovered when the cosmology of the template is different from that of the mocks, we generate a template computed with the Planck cosmology \citep{Planck2020}. We can  estimate the expected $\alpha$ by
\beq
\label{eq:alpha_Planck_shift}
\alpha=\frac{\left.\frac{r_{\mathrm{d}}}{D_{\mathrm{M}}}\right|_{\mathrm{MICE}}}{\left.\frac{r_{\mathrm{d}}}{D_{\mathrm{M}}}\right|_{\text {Planck }}}=0.959 ,
\eeq
where $r_{\mathrm{d}}$ is the sound horizon at the drag epoch and $D_{\mathrm{M}}$ is the comoving angular diameter distance evaluated at the effective redshift = 0.835  following DES Y3.  For MICE cosmology, $r_{\mathrm{d}}=153.4 \, \mathrm{Mpc}$ and $D_{\mathrm{M}}\left(z_{\text {eff }}\right) = 2959.7 \,  \mathrm{Mpc}$, while for Planck cosmology $r_{\mathrm{d}}=147.6  \, \mathrm{Mpc}$ and $D_{\mathrm{M}}\left(z_{\text {eff }}\right)=2967.0  \, \mathrm{Mpc}$.

\change{ Unlike other tests, when an alternative fiducial cosmology is assumed, the measurement of $ \xi_{\rm p} $ should be performed again as the distance has to be computed in the new cosmology. Furthermore, we mentioned that the Hubble parameter $h$ is  contained in the unit $\MpcOh$ and the distance does not explicitly depend on it.  
}

\change{  In Table \ref{tab:BAO_estimate_comparsion}, we compare the values of the $\alpha$ parameter obtained by the simple estimate using Eq.~\eqref{eq:alpha_Planck_shift} and the mock measurement results. In contrast, we also show similar results for $w$ from \citet{y3-baomain}. Interestingly, we find that the $ \xi_{\rm p} $ results are closer to the simple estimate than the ones from $w$. The deviation from the simple estimate could stem from the nonlinear correction and other systematic effects, and they affect the results from both templates in a similar manner. }


\begin{table}
  \caption{ Comparison of the $\alpha $ values obtained with the Planck and the MICE templates in fitting to the MICE cosmology data. The simple estimates using Eq.~\eqref{eq:alpha_Planck_shift} are shown in the first row, and the mock measurement in the second. As a reference, the mock measurements for $w$ from \citet{y3-baomain} are also shown in the last row.  }
  \label{tab:BAO_estimate_comparsion}
  \begin{tabular}{lccccccccc}
  \hline
  \hline
                                 &  Planck    &   MICE     \T\B  \\  \hline 
  Simple estimate                &  0.959              &   1               \\ \hline
  $\xi_{\rm p} $ mock results   &  0.958              &   1.001               \\ \hline
  $ w $ mock results         &  0.966              &   1.004                 \\ 
\hline \hline
  \end{tabular}
\end{table}

\begin{table*}
  \caption{ Comparison of the $\xip$ and $w$ BAO fit results, for which only the data in a single tomographic bin is used. }
  \label{tab:BAO_xip_vs_w}
  \begin{tabular}{lccccccccc}
  \hline
  \hline
  case &  $\langle \alpha \rangle$     &   $\sigma_{\rm std}$   &   $\sigma_{68}$      & $\langle \sigma_{\alpha} \rangle$    &  ${\rm frac.\,encl.}   \langle \alpha \rangle$   &  $\langle d_{\rm norm} \rangle$   & $\sigma_{d_{\rm norm}}$  &   ${\rm mean\,of\,mocks}$     &  $\langle \chi^2 \rangle / {\rm dof}$       \T\B  \\  \hline 
1, $\xip$          &  0.996   &  0.050  & 0.049  & 0.036   &  45\%    & 0.051      &  1.48    &  1.002 $\pm$ 0.039     &  44.3/29 (1.53)    \\ 
1, $w$             &  0.996   &  0.049  & 0.045  & 0.045   &  66\%    & -0.026      &  1.18    &  0.995  $\pm$ 0.049   &  17.4/17 (1.02)    \\
\hline
2, $\xip$          &  0.997   &  0.044  & 0.040  & 0.033   &  52\%    & -0.025      &  1.33    &  0.999 $\pm$ 0.037     &  44.2/29 (1.53)    \\ 
2, $w$             &  1.000   &  0.049  & 0.047  & 0.045   &  67\%    & -0.053      &  1.19    &  0.999 $\pm$ 0.049     &  16.3/17 (0.96)    \\
\hline
3, $\xip$          &  1.001   &  0.048  & 0.039  & 0.035   &  56\%    & -0.029      &  1.38    &  1.003 $\pm$ 0.035     &  44.8/29 (1.54)    \\ 
3, $w$             &  0.998   &  0.050  & 0.039  & 0.041   &  67\%    & -0.042      &  1.20    &  0.997 $\pm$ 0.043     &  17.6/17 (1.03)    \\
\hline
4, $\xip$          &  0.992   &  0.043  & 0.035  & 0.032   &  60\%    & -0.006      &  1.45    &  0.993 $\pm$ 0.034     &  39.1/29 (1.35)    \\ 
4, $w$             &  1.005   &  0.050  & 0.050  & 0.043   &  59\%    & -0.067      &  1.12    &  1.003 $\pm$ 0.043     &  17.2/17 (1.01)    \\
\hline
5, $\xip$          &  1.009   &  0.039  & 0.033  & 0.038   &  69\%    & -0.056      &  1.03    &  1.007 $\pm$ 0.039     &  43.7/29 (1.51)    \\ 
5, $w$             &  1.005   &  0.045  & 0.042  & 0.042   &  66\%    & -0.027      &  1.06    &  1.006 $\pm$ 0.047     &  18.9/17 (1.11)    \\
\hline \hline
  \end{tabular}
\end{table*}

We have shown the results obtained with the original Gaussian covariance. The $\rho $ modification prescription only  affects a small fraction of the mocks and the results are statistically similar to the original Gaussian case.   Moreover, although the COLA covariance suffers from the overlap issue, we also show its results for reference. Despite the overlap issue, the results, especially the measures of the spread and the error bar,  are similar to the default ones.  Because we are fitting the mocks with the same covariance derived from it, the $\chi^2 / \mathrm{ dof}=1.65 $ is smaller, but still it is much larger than 1. Overall, these suggest the consistency of our results across the covariance used.

The sequential maximum likelihood fitting is applied mostly for its efficiency and convenience. We have checked our results using MCMC, implemented by {\tt emcee } \citep{emcee2013}, in which all the parameters are fit simultanously. Similar to the conclusion for $w$ in \citet{Chan:2018gtc}, we find that the MCMC fit gives consistently similar results for $\xi_{\rm p}$.

Furthermore, we have considered splitting the data in the $\mu$ range [0,0.8] into three equal-$\mu$ bins and dividing the redshift range [0.6,1.1] into two sub-redshift bins: [0.6, 0.85] and [0.85, 1.1]. We have also included cross-$\mu$ bin or cross sub-redshift bin covariances. In both cases we find that the discrepancy between $\sigma_{\rm std }  $ and $\langle \sigma_{\alpha } \rangle $ increases. The issue is more serious for $N_{ \mu } = 3 $ case, and these invite  the interpretation that the discrepancy increases as the number of bins increases. We also see similar  trend in the smallest $ \Delta s $  bin case.   These could be attributed to the high covariance of the data as the correlated data are more susceptible to random fluctuations. By dividing the data into more bins, there are larger random fluctuations. On one hand, there are larger fluctuations in the best fit resulting in an increase in the spread of the distribution. On the other hand, the error bar derived from the local curvature of the likelihood function could be underestimated as the likelihood is less smooth and the curvature around the minimum is increased.

\change { In Table \ref{tab:BAO_mocktest}, we already compared the $\xip$ results in the photo-$z$ range [0.6,1.1] against the $w$ results obtained using the joint fit with five tomographic bins, each of bin width  $\Delta z_{\rm p} = 0.1 $. It is instructive to compare the performance of $\xip$ and $w$ for individual tomographic  bins, and the results are shown in Table \ref{tab:BAO_xip_vs_w}. In this test the intrinsic photo-$z$ distribution is comparable to the bin width. Unlike the joint-bin fit, there is no cross-bin information.  The difference between these statistics is that the data is further projected onto a sphere for $w$.  Because of the limitation in computing power, only 100 mocks are used in this comparison.  Overall, the results for $w$ are more stable across redshift bins. 
 Except for the first bin, the measures of the spread of the distribution are larger for  $w$, and this can be interpreted as information loss due to projection. However, we note that for $\xip $,   $\langle  \sigma_\alpha \rangle $ tends to be smaller than the measures of the spread by a larger amount. Besides, $\sigma_{d_{\rm norm} }$ is also more significantly larger than 1. These are caused by the increase in noise as the data size reduces and are consistent with the trends for $N_{\rm mu}  =3 $ and $N_z =2 $ in Table \ref{tab:BAO_mocktest}.  The $\chi^2 /{\rm dof }  $ for $\xip$ is still substantially larger than 1 although it is smaller than the joint fit case, while for $w$, it remains close to 1.   In summary, $\xip$ makes use of more information and hence potentially more constraining, but it is more susceptible to noise in the data.   }

\change{ Before closing this section, we would like to comment further on the dual roles of $\chi^2 $: the characterization of the goodness of fit and the model differentiation.  The $\chi^2 / {\rm dof} $ is often used to indicate the goodness of the fit \footnote{ Strictly speaking, the degrees of freedom is only defined for linear models as stressed in \citet{Andrae_etal_2010}, while the parameter $\alpha$ (and $B$) appears nonlinearly. Here, we followed the usual practice to assume that the degrees of freedom can be approximated defined for our model.  }.  We discussed previously that the $ \chi^2  / {\rm dof} $ is substantially larger than 1 although the model offers a good fit to the BAO. A separate question is if we can use the $\chi^2 $ value for model differentiation, especially for claiming the significance of the BAO detection. Although we may not be able to use the $\chi^2 $ distribution results, in principle we can use the mocks to answer the question that if there are no BAO signals in the data, how much the chance is that the BAO template  yields a  $\chi^2 $ value smaller than certain threshold. }

\section{Conclusions}
\label{sec:Conclusions}

There are numerous large-scale structure photometric surveys ongoing or upcoming. One of the important probes to extract cosmological information from these surveys is via the measurement of the BAO. A common way to measure BAO in photometric data is to divide the sample into tomographic bins and measure their auto (and cross) correlation.

It has been suggested to analyze the photometric data  as the spectroscopic ones through the 3D correlation function $\xi_{\rm p}$. Previous modeling was limited to a Gaussian photo-$z$ approximation. In this work, we generalize the modeling to accomodate general photo-$z$ distribution. This eliminates the possibility of biasing the results by assuming Gaussian photo-$z$ approximation.  To do so, we bin the general cross angular correlation between different redshift bins  $w_{ij} (\theta_k) $ using the variables $s$ and $\mu$ (or $s_\perp$ and $\mu$) like in usual 3D correlation function analysis.  Our approach highlights the connection between the angular correlation function and  3D correlation function.

To help understanding the impact of the photo-$z$ uncertainties on the correlation function, especially the BAO, we specialize the calculations to the Gaussian photo-$z$ case.   When $\sigma \gtrsim 0.02 $, the BAO peak in the correlation shows up at the true BAO scale only if the transverse scale of the separation, $s_{\perp}$, is used as the independent variable.    We show that this results from the interplay between the photo-$z$ distribution and the Jacobian of the transformation.  The latter formally diverges at $s_{\perp}$, and for  $\sigma \gtrsim 0.02 $, it dominates the integral, causing $\xi_{\rm p }$ to trace the underlying correlation function at the scale $ s_\perp $. Large-scale photometric surveys equipped with broadband filters such as DES typically have  $\sigma \gtrsim 0.02 $. In this case, it is favorable to adopt the parametrization  $\xi_{\rm p }(s_\perp, \mu )$, which effectively projects the 3D correlation function to the transverse direction. Unfortunately, this also means that $\xi_{\rm p } $ only probes the transverse BAO and cannot be used to directly constrain the Hubble parameter.

Our approach also enables us to derive the Gaussian covariance for $\xi_{\mathrm{p}}$ from the Gaussian covariance for the angular correlation function $w$.  Due to photo-$z$ mixing, the covariance of the $\xi_{\mathrm{p}}$ statistics has strong off-diagonal elements. This high correlation causes problems to the data fitting. One of the problems is that some of the best fit results are manifestly poor, but we find that  they can be resolved by suppressing the large eigenvalue modes in the covariance. Another issue is that the mean $\chi^{2} / {\rm dof}$ for the fit is large, even though the fit is apparently good.  While  this second issue cannot be easily resolved by adjusting the covariance, we find that the issue is primarily caused by the small eigenvalue modes and  they are not directly related to the BAO scale. We conclude that the BAO scale is well-fit by our model (Fig.~\ref{fig:Uy_NoBAO}).

We have used a set of dedicated DES Y3 mocks to test our pipeline. This set of mocks include realistic photo-$z$ distribution among other things. We find that the theory template is in good agreement with the mock measurement.  After verifying our methodology, we apply our pipeline to the BAO fit in the mock catalog. Overall we find that the results are consistent with the ones derived from  $w$ even with small improvements because  $\xi_{\rm p } $ makes use of the cross redshift bin correlation as well. However, the angular correlation function measurements are less correlated with  $\chi^{2} / {\rm dof} \sim 1$. As  $\xi_{\rm p}$ offers an alternative way to extract the BAO information, we shall apply it to the DES Y3 samples. This can serve as important cross check as different statistics have different sensitivities to the potential systematics.  Some preliminary results on $\xi_{\mathrm{p}}$ have been presented in \citet{y3-baomain} on the BAO samples, but a detailed quantitative analysis is still lacking. This is particularly interesting, as the $ D_{\rm M} $ measurement in \citet{y3-baomain} shows stimulating deviation from the Planck results. Measuring  $ D_{\rm M} $ using alternative statistics on various samples helps establish the results.


\section*{Acknowledgments}
We thank Fengji Hou for useful discussions.  K.C.C. acknowledges the support from the National Science Foundation of China under the grant 11873102, the science research grants from the China Manned Space Project with NO.CMS-CSST-2021-B01, and the Science and Technology Program of Guangzhou, China (No. 202002030360). S.A. is supported by the MICUES project, funded by the EU H2020 Marie Skłodowska-Curie Actions grant agreement no. 713366 (InterTalentum UAM) and ``EU-HORIZON-2020-776247 Enabling Weak Lensing Cosmology (EWC)''.

Funding for the DES Projects has been provided by the U.S. Department of Energy, the U.S. National Science Foundation, the Ministry of Science and Education of Spain, 
the Science and Technology Facilities Council of the United Kingdom, the Higher Education Funding Council for England, the National Center for Supercomputing 
Applications at the University of Illinois at Urbana-Champaign, the Kavli Institute of Cosmological Physics at the University of Chicago, 
the Center for Cosmology and Astro-Particle Physics at the Ohio State University,
the Mitchell Institute for Fundamental Physics and Astronomy at Texas A\&M University, Financiadora de Estudos e Projetos, 
Funda{\c c}{\~a}o Carlos Chagas Filho de Amparo {\`a} Pesquisa do Estado do Rio de Janeiro, Conselho Nacional de Desenvolvimento Cient{\'i}fico e Tecnol{\'o}gico and 
the Minist{\'e}rio da Ci{\^e}ncia, Tecnologia e Inova{\c c}{\~a}o, the Deutsche Forschungsgemeinschaft and the Collaborating Institutions in the Dark Energy Survey. 

The Collaborating Institutions are Argonne National Laboratory, the University of California at Santa Cruz, the University of Cambridge, Centro de Investigaciones Energ{\'e}ticas, 
Medioambientales y Tecnol{\'o}gicas-Madrid, the University of Chicago, University College London, the DES-Brazil Consortium, the University of Edinburgh, 
the Eidgen{\"o}ssische Technische Hochschule (ETH) Z{\"u}rich, 
Fermi National Accelerator Laboratory, the University of Illinois at Urbana-Champaign, the Institut de Ci{\`e}ncies de l'Espai (IEEC/CSIC), 
the Institut de F{\'i}sica d'Altes Energies, Lawrence Berkeley National Laboratory, the Ludwig-Maximilians Universit{\"a}t M{\"u}nchen and the associated Excellence Cluster Universe, 
the University of Michigan, NSF's NOIRLab, the University of Nottingham, The Ohio State University, the University of Pennsylvania, the University of Portsmouth, 
SLAC National Accelerator Laboratory, Stanford University, the University of Sussex, Texas A\&M University, and the OzDES Membership Consortium.

Based in part on observations at Cerro Tololo Inter-American Observatory at NSF's NOIRLab (NOIRLab Prop. ID 2012B-0001; PI: J. Frieman), which is managed by the Association of Universities for Research in Astronomy (AURA) under a cooperative agreement with the National Science Foundation.

The DES data management system is supported by the National Science Foundation under Grant Numbers AST-1138766 and AST-1536171.
The DES participants from Spanish institutions are partially supported by MICINN under grants ESP2017-89838, PGC2018-094773, PGC2018-102021, SEV-2016-0588, SEV-2016-0597, and MDM-2015-0509, some of which include ERDF funds from the European Union. IFAE is partially funded by the CERCA program of the Generalitat de Catalunya.
Research leading to these results has received funding from the European Research
Council under the European Union's Seventh Framework Program (FP7/2007-2013) including ERC grant agreements 240672, 291329, and 306478.
We  acknowledge support from the Brazilian Instituto Nacional de Ci\^encia
e Tecnologia (INCT) do e-Universo (CNPq grant 465376/2014-2).

This manuscript has been authored by Fermi Research Alliance, LLC under Contract No. DE-AC02-07CH11359 with the U.S. Department of Energy, Office of Science, Office of High Energy Physics.

\section*{Data availability}
The data underlying this article will be shared on reasonable request to the corresponding author.


\bibliographystyle{mnras}
\bibliography{references}

\appendix

\section{ Checks on the spectroscopic sample }

\label{sec:specz_check}

\begin{table*}
  \caption{ The BAO fit results for the spectroscopic sample. For each of the case shown, the first entry denotes the type of template, mock mean or theory, and the second one represents the type of covariance, mock or Gaussian covariance.   } 
  \label{tab:BAO_specz}
  \begin{tabular}{lccccccccc}
  \hline
  \hline
  Case &  $\langle \alpha \rangle$     &   $\sigma_{\rm std}$   &   $\sigma_{68}$      & $\langle \sigma_{\alpha} \rangle$    &  ${\rm frac.\,encl.}   \langle \alpha \rangle$   &  $\langle d_{\rm norm} \rangle$   & $\sigma_{d_{\rm norm}}$  &   ${\rm mean\,of\,mocks}$     &  $\langle \chi^2 \rangle / {\rm dof}$       \T\B  \\  \hline 
Mock, Mock               &  1.000   &  0.010  & 0.010   & 0.007     &  70\%     & -0.029      &  1.60    & 1.000   $\pm$ 0.009     & 24.0/29  (0.83)   \\
Mock,  Gaussian          &  1.000   &  0.010   & 0.011   &  0.005   &  52\%     & -0.022      & 1.43     & 1.001   $\pm$ 0.011     & 30.9/29  (1.07)    \\
Theory, Mock             &  1.001   &  0.013  & 0.014   &  0.011    &  74\%     & -0.005      &  1.25    &  1.001  $\pm$ 0.012     &  36.6/29 (1.26)    \\
Theory, Gaussian         &  1.002   &  0.014  & 0.014   &  0.007    &  44\%     & -0.023      &  1.38    &  1.002  $\pm$ 0.007     &  43.1/29 (1.49)    \\ 
\hline \hline
  \end{tabular}
\end{table*}

Although our goal is to consistently include photo-$z$ in our modeling, it is helpful to check our methodology on the redshift-space correlation function.

\begin{figure}
\centering
\includegraphics[width=\linewidth]{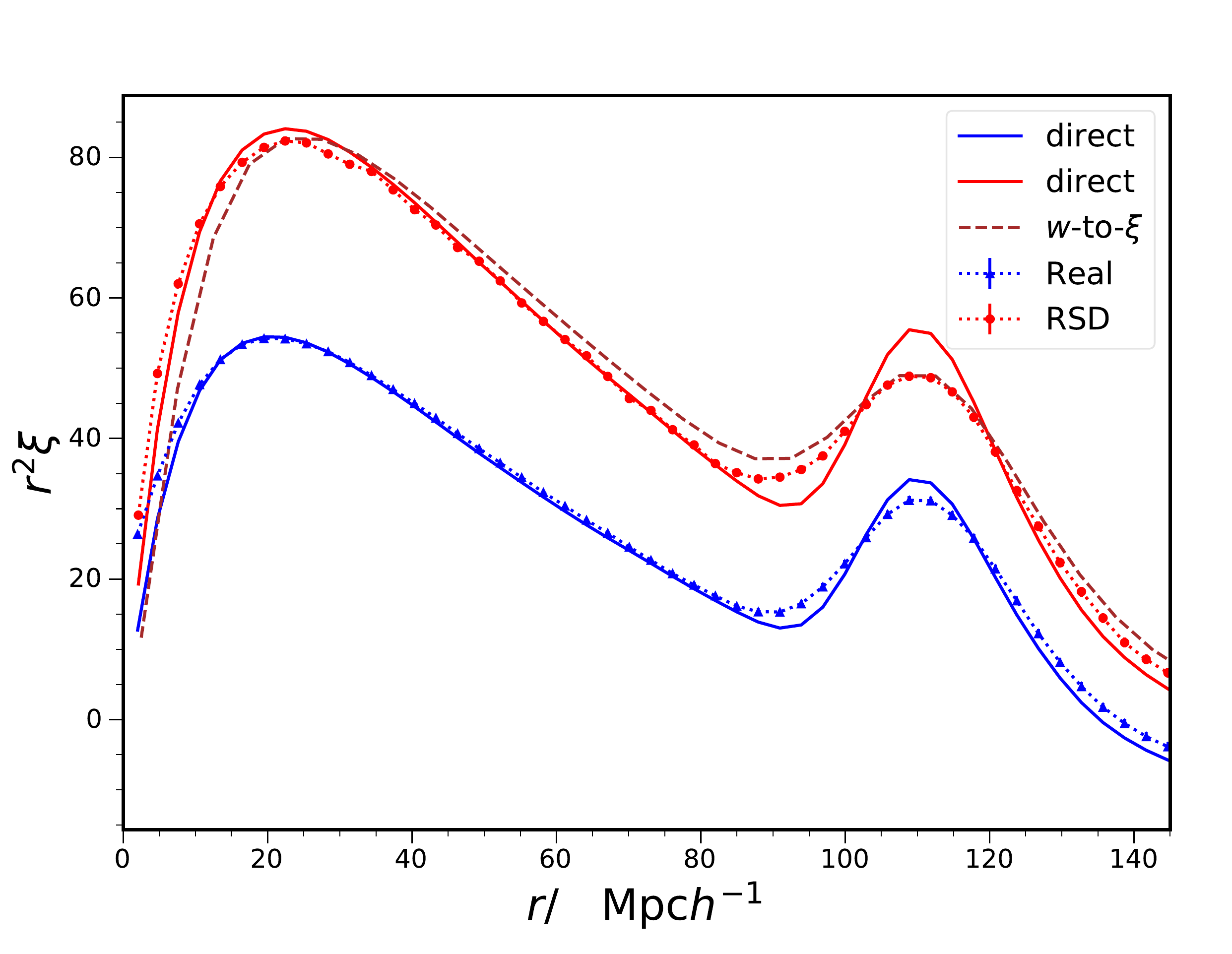}
\caption{ The real-space (blue, triangles) and redshift-space (red, circles) correlation measured from the ICE-COLA mocks. The standard linear templates obtained by direct computation (solid) are compared with the corresponding measurements. We also show the redshift-space template obtained by mapping $w$ to $\xi$ (brown, dashed).        }
\label{fig:xi_realz_rsdz_fit_neat}
\end{figure}

To test our pipeline, we use the real-space and redshift-space correlation function measured from the ICE-COLA mocks. In order to verify our template method, we also compute the standard linear redshift-space  3D correlation function template directly, which can be obtained by replacing  $\phi$ in Eq.~\eqref{eq:w_Kaiser_linbias} by a Dirac delta distribution. The real-space template can be retrieved by further setting the $f$-terms in Eq.~\eqref{eq:Al_Kaiser} to zero.   In Fig.~\ref{fig:xi_realz_rsdz_fit_neat},  we show the real-space correlation function measurement and the corresponding best-fit  template with the bias as the fitting parameter.  The real-space template is in good agreement with the measurements for $ r \gtrsim 20 \MpcOh $. Note that the BAO feature is stronger in the template as the linear power spectrum is used.   The best-fit effective bias is 1.63, which roughly corresponds to the bias value at the effective redshift 0.83 shown in Fig.~\ref{fig:bias_z_Nzbins} (or Fig.~\ref{fig:bias_photoz_specz} below).   Moreover, we compare the redshift-space correlation measurement with the corresponding template from the standard direct computation. The wedge correlation function is averaged over  $\mu$ in the same way as $\xip$ in the main text. We find that the agreement is mildly worse than the real-space result, and the BAO feature damping appears to be weaker than in the real-space case.  In the computation, we have used the best-fit effective bias parameter from the real-space result  so as to compare with the template obtained by mapping $w$ to $\xi$ below.   We could have used the bias parameter obtained by fitting to the redshift-space correlation and the agreement on small scales would be slightly improved.

\begin{figure}
\centering
\includegraphics[width=\linewidth]{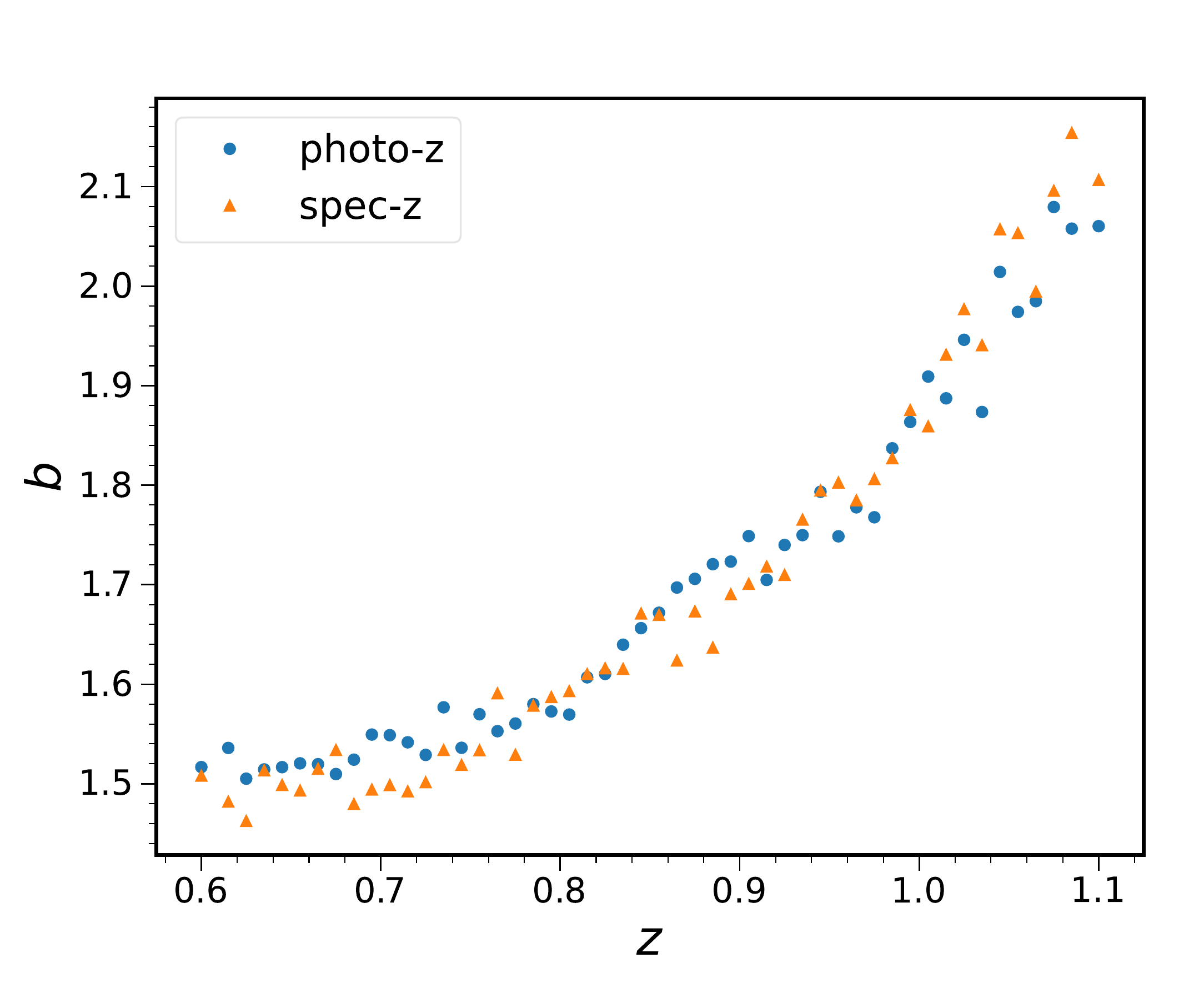}
\caption{ The best-fit bias obtained from the photometric sample (blue, circles) and spectroscopic sample (orange, triangles).  }
\label{fig:bias_photoz_specz}
\end{figure}

We also compute the spectroscopic template following the method of mapping $w$ to $\xi$ discussed in the main text.  Similar to the analysis for the photo-$z$ sample, we first measure the bias parameters of the redshift-space sample using the angular correlation function. In Fig.~\ref{fig:bias_photoz_specz}, we compare the bias parameters measured from the photo-$z$ angular correlation function and the spectroscopic angular correlation.  In both cases, we divide the data in the redshift range [0.6,1.1] into 50 equal redshift bins.  For the spectroscopic sample, the window function is a top-hat of width $ \Delta z= 0.01$.  We find that the recovered tomographic bin bias parameters are consistent with each other, and this demonstrates that fine photo-$z$ bin measurements can indeed recover the underlying intrinsic bias.

We go on to use the bias parameters to compute the 3D correlation. For the case of spectroscopic redshift, we compute the cross angular correlation function $ w_{ij}(\theta )$ with the the sampling width $\Delta z = 0.00333$ and $\Delta \theta = 0.02^{\circ} $. We find that finer sampling in $z$ does not improve the results.   Note that in this case  we have used the linear power spectrum with BAO damping.  The template  is in decent agreement with the measurement, but in the intermediate range 30 to 80 $\MpcOh$ it is slghtly worse than the direct computation result. The difference could be caused by the extra binning and averaging in the $w$ to $\xi$ mapping and these  may result in smoothing of the correaltion function. However, this is less of an issue in the photo-$z$ case as the intrinsic photo-$z$ mixing window is much stronger.




In Table \ref{tab:BAO_specz}, we show the BAO fit results for the spectroscopic samples. We have shown four scenarios with different template and covariance combinations. The template is either the theory one or the mock mean. The theory template is obtained by mapping $w$ to $\xi$. As there is small difference between the mock measurement and the theory, we also consider directly using the mean of the mock measurement as the template. In passing, in the main text this possiblity is not presented because we found that using the photo-$z$ $\xip$ mock mean as the template leads to a non-smooth likelihood.   Two covariances are considered, the mock covariance and the Gaussian one. The Gaussian covariance is computed in a similar way as the photo-$z$ case with 50 bins in redshift.

When both the template and the covariance are from mocks, it could be the best case scenario. But we still find that the error estimate is smaller than the spread of the distribution and $d_{\rm norm} $ deviates from Gaussianity significantly. This suggests that the likelihood is non-Gaussian. Moreover, the  $ \langle \chi^2 \rangle / {\rm dof} $ is small relative to 1. This could be caused by issues of the mocks such as the overlapping issue discussed in the main text.   When the mock covariance is replaced by the Gaussian one, $\langle \sigma_\alpha \rangle $ is further reduced and the fraction enclosing $ \langle \alpha \rangle $ is lowered. Interestingly, $ \langle \chi^2 \rangle / {\rm dof} $  is closest to unity in this case. When the theory template is used with the mock covariance, the spread of the best-fit $ \alpha $ and $\langle  \chi^2  \rangle / {\rm dof} $ increase.  These can be attributed to the difference between the template and the mock mean. Finally, when both the theory template and the Gaussian covariance are used, the  $ \langle \chi^2 \rangle / {\rm dof} $ is the highest. We note that it is still smaller than the photo-$z$ case by 0.5.





\bsp	
\label{lastpage}
\end{document}